# Focal-Plane Extreme Adaptive Optics for Deep Super Resolution Imaging and Nanolithography

Jinxin Liu[1,2,†], Jiahao Niu[1,2,†], Runxin Zeng[1,2,†], Kaining Yang[1,2,†], Ziyu Li[3,†], Xiaodong Guan[2,4], Jing Ren[2], Kebin Shi[2,3], Yue Wang[2,3,*], Shan Zhao[1,2,*], Yujie Sun[1,2,4,*], He Sun[1,2,*]

*1 College of Future Technology, Peking University, Beijing, China*
*2 National Biomedical Imaging Center, Peking University, Beijing, China*
*3 State Key Laboratory for Mesoscopic Physics and Frontiers Science Center for Nano-optoelectronics, School of Physics, Peking University, Beijing, China*
*4 School of Life Sciences, Peking University, Beijing, China*

† Contributed equally to this work: Jinxin Liu, Jiahao Niu, Runxin Zeng, Kaining Yang, Ziyu Li

* Corresponding authors: Y. Wang (yuew@pku.edu.cn), S. Zhao (shan.zhao@pku.edu.cn), Y. Sun (sun_yujie@pku.edu.cn) and H. Sun (hesun@pku.edu.cn)

## Abstract

Super resolution optical imaging and nanolithography enable nanoscale observation and fabrication, but high order, spatially varying phase and amplitude aberrations in optically heterogeneous media severely limit their performance at depth[1-7]. Here we introduce focal-plane extreme adaptive optics (FP-ExAO), which moves beyond conventional phase conjugate correction by directly sensing and controlling the focal-plane complex field. This terminal field captures aberrations accumulated throughout the optical system. A state space formulation directly links deformable mirror commands to the focal field, avoiding explicit plane by plane Fourier optics modelling, while information theoretically optimal probing and closed-form control enable rapid field estimation and correction. Cascaded deformable mirrors further enable FP-ExAO to correct coupled phase-amplitude and spatially varying aberrations.

In three-dimensional (3D) single molecule localization microscopy (SMLM), FP-ExAO enables imaging to 160 µm depth in brain tissue and beyond 60 µm in kidney tissue, while maintaining approximately 18 nm localization precision across a 75 µm × 75 µm corrected field, with a 16-fold larger field of view and more than tenfold faster correction than conventional AO methods. These capabilities meet the imaging requirements for 3D deep tissue optical connectome reconstruction and quantitative super resolution pathology. In stimulated emission depletion (STED) nanolithography, FP-ExAO reduces the minimum written linewidth from 251.1 nm without correction to 88.8 nm at a depth of 160 µm in photoresist. FP-ExAO establishes a new adaptive optics paradigm for deep nanoscale optical readout and writing.

## Introduction

Super resolution optical imaging and nanolithography enable nanoscale observation and fabrication, but their performance deteriorates rapidly in optically heterogeneous 3D media[1-5]. In biological tissues and photoresists, refractive index heterogeneity introduces aberrations throughout the optical path rather than at a single well-defined plane. As light propagates deeper, these distortions become increasingly high order and spatially varying, with coupled phase and amplitude components. These effects are particularly severe in super resolution systems, where precisely structured optical fields are often generated using diffractive optical elements (DOEs). Small aberrations can disrupt the interference patterns required for super resolution, while imperfections in the DOEs themselves introduce additional system errors. Consequently, maintaining nanoscale performance deep inside heterogeneous media requires correcting aberrations far more complex than those encountered in shallow or homogeneous samples[6,7].

Adaptive optics (AO) can mitigate optical aberrations, but existing approaches remain poorly suited to these complex 3D distortions[8,9]. Sensor-based AO directly measures wavefronts but requires additional hardware and guide stars, which are often impractical in photon limited fluorescence imaging or at the writing plane in nanolithography[10-12]. Sensorless AO avoids dedicated wavefront sensing but relies on task specific image metrics and repeated perturbations, making high order correction slow and fragile[13-16]. Phase diversity AO infers aberrations from intensity measurements through an explicit Fourier optics model, typically parameterized in a pupil or conjugate plane[17-18]. Extending such models to distributed, spatially varying 3D distortions is computationally expensive and vulnerable to model mismatch. Computational AO algorithmically compensates aberrations after detection, but cannot recover photons lost before detection or physically correct the optical fields used for fabrication[19]. Tissue clearing can reduce optical heterogeneity, but typically requires extensive sample processing and is not suited to in situ correction or optical fabrication[20]. These limitations make it difficult for existing AO methods to rapidly and accurately correct the high order, spatially varying, phase–amplitude distortions encountered in deep super resolution imaging and writing.

Here we introduce focal-plane extreme adaptive optics (FP-ExAO), which directly senses and controls the task relevant complex electric field in the focal plane, using this terminal field to capture the cumulative effects of aberrations throughout the optical system. Inspired by extreme adaptive optics for high contrast astronomical imaging[21-22], FP-ExAO formulates the end-to-end response from deformable mirror commands to the terminal field using a state space model, avoiding explicit pupil wavefront assumption and plane by plane Fourier optics modeling. Building on information theory and stochastic optimal control theory, we develop an optimal deformable mirror probing policy that introduces controlled intensity variations for efficient complex field estimation, together with a closed-form controller that drives the estimated field towards a task defined target. Because the model directly connects multiple wavefront correctors to the terminal field, the correctors need not be restricted to a single pupil or prescribed conjugate plane. This allows their combined degrees of freedom to be used cooperatively; in our implementation, cascaded deformable mirrors enable rapid correction of high order, spatially varying phase and amplitude distortions.

We demonstrate FP-ExAO across two distinct super resolution tasks: SMLM and STED nanolithography. In fluorescent bead experiments with spatially varying aberrations, FP-ExAO corrects aberrations more than 10 times faster than metric-based and phase diversity AO, while expanding the corrected field of view by 16-fold. In 3D SMLM, it enables imaging of TOM20-labelled mitochondria in COS-7 cells, GFP- and GFAP-labelled neuron axonal and astrocytic structures in brain tissue, and nephrin-labelled glomerular structures in kidney tissue. FP-ExAO reaches imaging depths of 160 μm in brain tissue and more than 60 μm in kidney tissue over a 75 μm × 75 μm corrected field of view, while maintaining approximately 18 nm localization precision. At depths beyond 150 μm in brain tissue, it resolves dendritic spine necks and axially oriented synaptic contacts, establishing the nanoscale imaging capability needed for optical connectome reconstruction in unexpanded deep tissue. In kidney tissue, the method resolves nephrin-positive slit diaphragm structures and quantifies morphological differences between healthy and pathological samples (e.g., focal segmental glomerulosclerosis, FSGS). In STED nanolithography, FP-ExAO enables sub-100 nm writing at a depth of 160 μm in photoresist, reducing the isolated line width from 251.1 nm without correction to 88.8 nm and improving the fidelity of dense gratings and complex patterns. These results establish focal-plane complex field sensing and control as a general extreme AO framework for preserving nanoscale optical performance in heterogeneous 3D media.

## Results

## FP-ExAO senses and controls the terminal complex field

Adaptive optics (AO) uses physical wavefront correctors, typically deformable mirrors (DMs) or spatial light modulators (SLMs), to shape the optical field for imaging or writing. An optical system containing multiple DM correctors can be modeled by a sequence of Fourier and Fresnel propagation operators and DM phase modulations[23,24]. For the dual DM configuration considered here, this propagation can be written schematically as:

$$E_f = \mathcal{F}\{E_{ab} \cdot exp\{i\varphi_1\} \cdot H_z\{exp\{i\varphi_2\}\}\} \quad (1.1)$$

where $E_{ab}$ denotes the aberrated field entering the correction system at the objective back pupil, $E_f$ is the terminal focal plane field, $\varphi_1$ and $\varphi_2$ are the phase modulations introduced by the DMs located at the pupil plane and an axially separated out of pupil plane, respectively. $\mathcal{F}$ denotes lens mediated Fourier propagation and $H_z$ denotes Fresnel propagation over distance $z$. AO systems compute the DM modulations so that the resulting point spread function (PSF) approaches the field distribution required by the imaging or writing task. This defines a nonlinear optical model and a non convex control problem.

Conventional AO addresses this problem by retrieving the aberrated wavefront at an assumed pupil or conjugate plane, either directly with a wavefront sensor or indirectly from intensity variations induced by controlled perturbations, and then applying phase conjugate compensation. In deep heterogeneous media and optical systems containing diffractive elements, however, distortions accumulate across multiple planes during propagation, leading to coupling between phase and amplitude. The resulting aberrations are high order, spatially varying and difficult to represent as phase only errors at any single plane. Inaccuracies in propagation distance, optical alignment, element fabrication and specimen structure further produce model mismatch. These effects make phase conjugation increasingly ineffective for deep super resolution imaging and nanolithography.

FP-ExAO instead aims to directly regulate the focal plane complex electric field (Fig. 1a). Rather than explicitly modeling how aberrations at different planes propagate through the optical system, we linearize the complete end-to-end optical response to the DM commands around the current operating point. For a small DM perturbation $\Delta u_k$, the phase response can be approximated by a first order Taylor expansion:

$$e^{i\frac{4\pi}{\lambda}\cdot\phi(\Delta u_k)} = 1 + i\frac{4\pi}{\lambda}\cdot\phi(\Delta u_k) + o(\Delta {u_k}^2) \quad (1.2)$$

This leads to the linearized focal-field response:

$$E_{f,k} = E_{f,k-1} + G\Delta u_k \tag{1.3}$$

or, in a real-valued state space representation:

$$x_{f,k} = x_{f,k-1} + G_R \Delta u_k \tag{1.4}$$

where $x_{f,k} \triangleq [\Re\{E_{f,k}\}, \Im\{E_{f,k}\}]$, $\Delta u_k$ contains the voltage increments applied to all DMs, $G$ and $G_R$ are the complex and real-valued Jacobian matrices that approximate the end-to-end optical response. Importantly, the Jacobian matrices directly capture how all physical correctors jointly perturb the terminal complex field, including inter DM propagation and other optical effects. These response matrices are obtained through system calibration and characterize the local input–output relationship between DM commands and focal-field variations (Methods; Supplementary note 1.5). FP-ExAO therefore does not require aberrations to be explicitly assigned to individual optical planes or propagated through a detailed plane by plane Fourier optics model.

We next estimate this terminal complex field directly from focal-plane intensity measurements. FP-ExAO applies symmetric pairs of DM probes $\pm u_k^{p,m}$, which produce:

$$I_{f,k}^{m+} = \left|E_{f,k} + G u_k^{p,m}\right|^2 + n_{f,k}^{+}, I_{f,k}^{m-} = \left|E_{f,k} - G u_k^{p,m}\right|^2 + n_{f,k}^{-} \tag{1.5}$$

where $G u_k^{p,M}$ is the known focal-plane response to the probe and $n_{f,k}$ is measurement noise. Subtracting each pair eliminates the quadratic terms:

$$\begin{bmatrix} \Delta I_k^1 \\ \vdots \\ \Delta I_k^M \end{bmatrix} = \begin{bmatrix} I_{f,k}^{1+} - I_{f,k}^{1-} \\ \vdots \\ I_{f,k}^{M+} - I_{f,k}^{M-} \end{bmatrix} = 4\Re\left\{\begin{bmatrix} (G u_k^{p,1})^* \circ E_{f,k} \\ \vdots \\ (G u_k^{p,M})^* \circ E_{f,k} \end{bmatrix}\right\} + n_{f,k} \tag{1.6}$$

Each probe pair therefore measures the projection of the unknown complex field along a known complex direction (Fig.1d). Stacking $M$ pairs give the linear observation model $\delta_k = A x_{f,k} + n_{f,k}$, from which the field can be estimated in a least squared form:

$$\hat{x}_{f,k} = (A^T A)^{-1} A^T \delta_k \tag{1.7}$$

Because the observation model is separable by focal plane pixel, the full complex field can be recovered very efficiently through parallel pixel wise regression. Carefully designed probe shapes can recover the focal-plane complex field accurately from only a few pairs of intensity measurements, thereby accelerating adaptive optics correction. Information theoretic analysis shows that optimal probes should produce approximately uniform focal-plane perturbations while sampling complementary complex field directions near quadrature[25]. We realize these conditions using two-dimensional (2D) sinc function probes for broad focal-plane coverage and Hilbert transform pairs for approximately orthogonal complex responses (Fig. 1c, d; Supplementary Fig. 4).

Once the terminal field $\hat{x}_{f,k}$ is estimated, the state space model directly provides the optimal DM correction. Rather than reconstructing a phase aberration and applying phase conjugation, FP-ExAO solves an optimal control problem:

$$\Delta u_k = \underset{\Delta u}{\operatorname{argmin}} \left\| \sqrt{W} (\hat{x}_{f,k} + G_R \Delta u) \right\|_2^2 + \sqrt{\alpha} \|\Delta u\|_2^2 \qquad (1.8)$$

where $W$ selects and weights the task relevant focal-plane regions and $\alpha$ regularizes the command amplitude. In SMLM, for example, the controller suppresses optical energy outside the desired PSF core. Because the objective is quadratic under the locally linear model, the optimal command has the closed-form solution:

$$\Delta u_k = -( {G_R}^T W \, G_R + \alpha I)^{-1} \, {G_R}^T W \hat{x}_{f,k} \qquad (1.9)$$

This eliminates the iterative metric searches used in conventional sensorless AO and allows rapid closed loop focal field regulation. Because FP-ExAO jointly computes the optimal commands for all wavefront correctors, it enables cooperative phase-amplitude correction without assigning individual correctors to specific aberration planes. For spatially varying aberrations, the same framework yields a closed-form command that simultaneously optimizes the PSF across multiple isoplanatic regions, providing coordinated field wide correction.

**FP-ExAO rapidly corrects high order and field-dependent aberrations**

We first validated FP-ExAO in a microscope with field-dependent aberrations, using spatially distributed fluorescent beads as a large field imaging target. The system used a cascaded dual DM adaptive optics architecture, containing a 140 actuator DM placed at the pupil plane and a larger 952 actuator DM positioned 146 mm downstream. This system with cascaded DMs provides control degrees of freedom across a 75 μm × 75 μm field, enabling correction of both field-shared and field-dependent aberrations (Methods; Extended Fig. 7). We divided the field into 5 × 5 isoplanatic regions for spatially resolved sensing and control.

We first tested the focal-plane probing strategy. Under a fixed number of measurements, the information optimized 2D sinc probes produced more accurate field estimates and higher corrected Strehl ratios than Zernike or random probes (Fig. 2a). Within the sinc probe family, Hilbert transform pairs that generated approximately 90° angular separation in the complex plane consistently outperformed 30° and 45° probe separations, confirming the benefit of near orthogonal sampling of the two field quadratures. Increasing the number of probe pairs from two to four further improved correction, and four pairs were therefore used in subsequent experiments (Supplementary Fig. 4).

We next compared FP-ExAO with metric-based sensorless AO and phase diversity AO under low and high order phase aberrations and coupled phase-amplitude distortions. Across both simulations and microscope measurements, FP-ExAO restored narrower PSFs and consistently higher Strehl ratios than the baseline methods (Fig. 2b–d; Extended Fig. 4; Supplementary Fig. 9). On spatially distributed fluorescent beads, FP-ExAO reduced correction time by approximately tenfold while achieving two- to threefold higher corrected Strehl ratios. Its advantage persisted under reduced signal to background and signal to noise ratios and during repeated or abrupt changes in aberration (Supplementary Figs. 8 and 10).

Finally, we tested spatially varying aberrations across the 5 × 5 isoplanatic regions. Metric-based and phase diversity AO improved the central regions but left the edge PSFs distorted. In contrast, FP-ExAO maintained substantially higher and more uniform Strehl ratios across the full field (Fig. 2e, f; Extended Fig. 5; Supplementary Fig. 7).

**FP-ExAO enables large field three dimensional TOM20 SMLM**

We next demonstrated the method in 3D SMLM under stochastic single molecule blinking. 3D localization was enabled by astigmatic PSF engineering using the pupil plane DM (Methods)[26,27]. We used TOM20-labelled outer mitochondrial membranes in COS-7 cells as the biological sample and introduced spatially varying aberrations with the DMs. Compared with fluorescent beads, single molecule signals are weaker and emitters blink stochastically, making focal field estimation from individual frames impractical. For each positive and negative probe state, we therefore localized and cropped blinking PSFs within each isoplanatic region and averaged them to form a representative probe response for focal field estimation and control (Fig. 3a; Supplementary Fig. 3). This procedure does not require the same molecule to remain emissive throughout the probe sequence and enables field-dependent correction over a large field of view.

Across a 75 µm × 75 µm field, we compared NoAO, FP-ExAO with a single pupil plane

DM, and FP-ExAO with cascaded DMs. Both configurations employed the FP-ExAO framework for focal-field sensing and control. FP-ExAO with cascaded DMs recovered mitochondrial outer membranes with greater structural continuity, more clearly resolved hollow profiles and better defined XZ cross sections than either single DM correction or NoAO (Fig. 3b–j, Supplementary Fig. 16a-e). Correction also increased detected photon numbers and localizations per frame while improving the axial localization distribution (Fig. 3k, Supplementary Fig. 16f-h). With FP-ExAO using cascaded DMs, localization precision reached 14 nm laterally and 53 nm axially, and Fourier ring correlation (FRC) resolution improved from 127.1 nm without AO to 66.7 nm (Extended Fig. 6a). Across five additional blinking datasets, FP-ExAO also produced more tightly confined PSFs than metric-based and phase diversity AO (Supplementary Fig. 15). These results demonstrate that FP-ExAO remains effective under weak, stochastic SMLM signals, improving photon efficiency, localization precision and 3D structural fidelity.

**FP-ExAO resolves deep tissue neuronal nanostructures for 3D circuit mapping**

We next applied FP-ExAO to tissue induced aberrations in mouse brain, demonstrating that nanoscale neuronal structures could be resolved in three dimensions over a large field at depths beyond 150 µm. In Thy1-EGFP brain slices, FP-ExAO enabled 3D SMLM imaging at depths of 144.3–161.7 µm across a 75 µm × 75 µm corrected field of view (Fig. 4a, h, i). Compared with NoAO, correction substantially increased localization recovery and structural continuity and more clearly separated neuronal processes along the optical axis (Fig. 4b-g, j-o). Across three representative regions, lateral localization precision remained approximately 16–17 nm, while FRC resolution improved by 1.6–1.9 fold, from 75.5 to 40.7 nm, 112.1 to 71.3 nm and 103.1 to 58.8 nm, respectively (Extended Fig. 6b-d). The corrected field was approximately 16-fold larger in area than previous deep AO-SMLM fields confined to a single 15–20 µm wide isoplanatic region[28-32].

FP-ExAO revealed neuronal structures that were poorly resolved without correction. At a depth of 161.7 µm, FP-ExAO resolved the 3D arrangement of neighbouring neuronal processes, distinguishing structures that overlapped laterally but were separated axially (Fig. 4d–g). At depths of 144.3 and 156.2 µm, FP-ExAO resolved dendritic spines with neck widths of approximately 110–130 nm and head widths of approximately 300–600 nm (Fig. 4j–o), whereas several corresponding structures were poorly discernible without AO. We observed similar improvements in GFAP-labelled astrocytic fibres at 155.3 μm depth, with finer and more continuous structures and an improvement in FRC resolution from 74.6 to 60.8 nm (Extended Figs. 3 and 6h). These results demonstrate that FP-ExAO reaches the 3D nanoscale resolution needed for morphology-based optical connectome in deep tissue[33-35].

## FP-ExAO enables deep tissue super resolution 3D pathology in kidney

We next applied FP-ExAO to nephrin-labelled mouse kidney tissue, testing whether 3D SMLM could resolve nanoscale glomerular structures under deep tissue induced, spatially varying aberrations. Nephrin delineates the slit diaphragm network between interdigitating podocyte foot processes[36-40]. At a depth of 63.4 μm, FP-ExAO preserved continuous, meandering nephrin-positive structures across both the center and edges of the field, whereas pupil only correction degraded towards the field periphery and NoAO substantially reduced structural visibility and continuity (Fig. 5a–f). FP-ExAO achieved a lateral localization precision of 16.3 nm and improved FRC resolution from 61.7 nm without AO to 49.2 nm (Extended Fig. 6e). The resolved podocyte foot process (PFP) features were predominantly 150–300 nm wide (Fig. 5g, h). 3D reconstructions at depths of approximately 20 μm further revealed clearer nephrin organization and XZ structure after correction, confirming improved lateral and axial structural fidelity (Extended Fig. 1).

We examined whether this imaging capability could support quantitative super resolution pathology. Healthy glomeruli imaged with FP-ExAO showed dense, continuous nephrin-positive slit diaphragm networks between interdigitating foot processes and intact contours along capillary loops (Fig. 5i). In contrast, these structures progressively deteriorated with disease severity in focal segmental glomerulosclerosis (FSGS) samples. Mild FSGS showed reduced and fragmented nephrin-positive meanders with increasingly discontinuous capillary loop contours, whereas severe FSGS exhibited extensive loss of recognizable slit diaphragm organization (Fig. 5j,k; Extended Fig. 2). Quantitative analysis showed corresponding decreases in Slit diaphragm density (SD) and nephrin continuity ratio (NCR) and an increase in gap density from healthy to mild and severe FSGS (Fig. 5l). Notably, healthy tissue without AO correction closely resembled mild FSGS in the morphological readouts (Fig. 5i(ii-iii), 5j, 5l(ii-iii)), indicating that uncorrected tissue aberrations can obscure disease specific morphology and compromise nanoscale pathology assessment.

## FP-ExAO enables deep STED nanolithography

Finally, we tested whether FP-ExAO could extend from super resolution imaging to nanoscale optical writing. We integrated FP-ExAO into a STED direct laser writing system, in which a 780 nm excitation beam induced two-photon absorption and a co-aligned 532 nm depletion beam were focused into the photoresist[1,4,5,41-43]. In contrast to SMLM, where FP-ExAO controls the detection PSF, here the controller regulates the depletion field towards an ideal doughnut-shaped focus at the two-photon excitation plane, suppressing residual intensity at both the central null and outside the target annulus[1,44-46]. Fabricated structures were characterized by scanning electron microscopy (SEM).

At a writing depth of 160 µm, FP-ExAO reduced the isolated linewidth to 88.8 nm, compared with 128.4 nm using metrics-based AO and 251.1 nm without correction, enabling sub-100 nm fabrication deep inside the photoresist (Fig. 6a). For densely spaced gratings, FP-ExAO produced 121.8 nm lines separated by 64.8 nm gaps, compared with 319.1 nm lines and 261.7 nm gaps without correction (Fig. 6b, c). The broader linewidths in dense gratings than in isolated features were consistent with proximity dependent photochemical effects between neighbouring writing trajectories.

We fabricated complex patterns at the same 160 µm to assess structural fidelity beyond isolated lines. In the plum blossom pattern, FP-ExAO preserved thin petal boundaries, resolved densely packed stamens and pistil at the flower center and maintained distinct contours along peripheral branches, whereas these structures broadened or merged without correction (Fig. 6d, f, g, j). In the Temple of Heaven pattern, FP-ExAO preserved detailed architectural outlines, windows and step structures, that were substantially blurred under NoAO conditions (Fig. 6e, h, i, k). FP-ExAO enabled finer feature sizes and higher structural fidelity in deep STED nanolithography.

## Discussion

FP-ExAO reframes adaptive optics as direct regulation of the focal plane complex field. Conventional AO generally retrieves aberrations at a pupil or conjugate plane and applies phase conjugate correction, an approach that becomes incomplete when distortions are distributed through heterogeneous 3D media, vary across the field and couple phase and amplitude. FP-ExAO instead uses the terminal complex field to represent their cumulative optical effect and directly controls this state towards a task defined target. Building on focal-plane electric field control developed for astronomical extreme AO, our framework extends this principle to spatially varying aberrations, stochastic fluorescence signals and nanoscale optical writing, while information optimal probing and closed-form control enable efficient terminal field modulation.

FP-ExAO extends nanoscale optical readout and writing deep into heterogeneous 3D media. In brain tissue, FP-ExAO maintained 3D SMLM performance beyond 150 μm and resolved dendritic spine morphology and the axial organization of neighbouring neuronal processes, achieving the nanoscale structural resolution required for optical connectome reconstruction. In kidney tissue, deep SMLM resolved nephrin-positive slit diaphragm organization and disease associated morphological changes. Healthy tissue imaged without AO partially resembled mild pathology in quantitative readouts, showing that tissue aberrations can themselves introduce biological measurement bias. In STED nanolithography, the same control principle extended high fidelity sub-100 nm fabrication to depths of 160 μm.

The present implementation still has limitations. The state space model relies on local linearity and can therefore degrade under large optical drift or abrupt changes in aberration. However, because the operating point evolves gradually during closed-loop correction, the local system response can be updated through online system identification and Jacobian refinement. Focal-field estimation also requires sufficient information about the local optical response, which is naturally available from point like emitters in SMLM but may require alternative guide signals in other modalities. In addition, increasingly fine spatial correction requires more isoplanatic regions, creating a trade-off between field of view, measurement burden and correction bandwidth. Although we used at most two cascaded deformable mirrors, FP-ExAO is architecture agnostic and can in principle coordinate an arbitrary number of optical actuators. Cascading additional deformable mirrors may enable Strehl ratios above 0.9 under 3D aberrations and ultimately help extend multiphoton brain imaging through the intact skull[47-50] to depths approaching 2 mm. Even in cleared thick tissues, residual aberrations may remain, and FP-ExAO could provide complementary correction to further improve image quality.

More broadly, FP-ExAO defines a new design principle for complex adaptive optical systems: rather than explicitly reconstructing internal distortions that are difficult to observe or model, an instrument can control the end-to-end response from its optical actuators to the task relevant output. This brings optical sensing, actuation and system design into a common closed loop framework, providing a general strategy for operating complex optical instruments near their physical limits.

## Reference

1. Hell, S. W. & Wichmann, J. Breaking the diffraction resolution limit by stimulated emission: stimulated-emission-depletion fluorescence microscopy. Opt. Lett. 19, 780–782 (1994).
2. Betzig, E. et al. Imaging intracellular fluorescent proteins at nanometer resolution. Science 313, 1642–1645 (2006).
3. Rust, M. J., Bates, M. & Zhuang, X. Sub-diffraction-limit imaging by stochastic optical reconstruction microscopy (STORM). Nat. Methods 3, 793–796 (2006).
4. Scott, T. F., Kowalski, B. A., Sullivan, A. C., Bowman, C. N. & McLeod, R. R. Two-color single-photon photoinitiation and photoinhibition for subdiffraction photolithography. Science 324, 913–917 (2009).
5. Gan, Z., Cao, Y., Evans, R. A. & Gu, M. Three-dimensional deep sub-diffraction optical beam lithography with 9 nm feature size. Nat. Commun. 4, 2061 (2013).
6. Booth, M. J. Adaptive optical microscopy: the ongoing quest for a perfect image. Light Sci. Appl. 3, e165 (2014).
7. Ji, N. Adaptive optical fluorescence microscopy. Nat. Methods 14, 374–380 (2017).
8. Zheng, W. et al. Adaptive optics improves multiphoton super-resolution imaging. Nat. Methods 14, 869–872 (2017).
9. Qiao, C. et al. Fast-adaptive super-resolution lattice light-sheet microscopy for rapid, long-term, near-isotropic subcellular imaging. Nat. Methods 22, 1059–1069 (2025).
10. Ji, N., Milkie, D. E. & Betzig, E. Adaptive optics via pupil segmentation for high-resolution imaging in biological tissues. Nat. Methods 7, 141–147 (2010).
11. Wang, K. et al. Rapid adaptive optical recovery of optimal resolution over large volumes. Nat. Methods 11, 625–628 (2014).
12. Wang, K. et al. Direct wavefront sensing for high-resolution in vivo imaging in scattering tissue. Nat. Commun. 6, 7276 (2015).
13. Débarre, D., Booth, M. J. & Wilson, T. Image based adaptive optics through optimisation of low spatial frequencies. Opt. Express 15, 8176–8190 (2007).
14. Siemons, M. E., Hanemaaijer, N. A. K., Kole, M. H. P. & Kapitein, L. C. Robust adaptive optics for localization microscopy deep in complex tissue. Nat. Commun. 12, 3407 (2021).
15. Žurauskas, M. et al. IsoSense: frequency enhanced sensorless adaptive optics through structured illumination. Optica 6, 370–379 (2019).
16. Hu, Q. et al. A universal framework for microscope sensorless adaptive optics:

Generalized aberration representations. APL Photonics 5, 100801 (2020).
17. Kner, P. Phase diversity for three-dimensional imaging. J. Opt. Soc. Am. A 30, 1980–1987 (2013).
18. Johnson, C. et al. Phase diversity-based wavefront sensing for fluorescence microscopy. Optica 11, 806–820 (2024).
19. Wu, J. et al. Iterative tomography with digital adaptive optics permits hour-long intravital observation of 3D subcellular dynamics at millisecond scale. Cell 184, 3318–3332.e17 (2021).
20. Chung, K. et al. Structural and molecular interrogation of intact biological systems. Nature 497, 332–337 (2013).
21. Sun, H., Kasdin, N. J. & Vanderbei, R. Identification and adaptive control of a high-contrast focal plane wavefront correction system. J. Astron. Telesc. Instrum. Syst. 4, 049006 (2018).
22. Pueyo, L. et al. Optimal dark hole generation via two deformable mirrors with stroke minimization. Appl. Opt. 48, 6296–6312 (2009).
23. Simmonds, R. D. & Booth, M. J. Modelling of multi-conjugate adaptive optics for spatially variant aberrations in microscopy. J. Opt. 15, 094010 (2013).
24. Wu, T.-W. & Cui, M. Numerical study of multi-conjugate large area wavefront correction for deep tissue microscopy. Opt. Express 23, 7463–7470 (2015).
25. Sun, H., Kasdin, N. J. & Vanderbei, R. Efficient wavefront sensing for space-based adaptive optics. J. Astron. Telesc. Instrum. Syst. 6, 019001 (2020).
26. Huang, B., Wang, W., Bates, M. & Zhuang, X. Three-dimensional super-resolution imaging by stochastic optical reconstruction microscopy. Science 319, 810–813 (2008).
27. Huang, B., Jones, S. A., Brandenburg, B. & Zhuang, X. Whole-cell 3D STORM reveals interactions between cellular structures with nanometer-scale resolution. Nat. Methods 5, 1047–1052 (2008).
28. Burke, D., Patton, B., Huang, F., Bewersdorf, J. & Booth, M. J. Adaptive optics correction of specimen-induced aberrations in single-molecule switching microscopy. Optica 2, 177–185 (2015).
29. Mlodzianoski, M. J. et al. Active PSF shaping and adaptive optics enable volumetric localization microscopy through brain sections. Nat. Methods 15, 583–586 (2018).
30. Zhang, P. et al. Deep learning-driven adaptive optics for single-molecule localization microscopy. Nat. Methods 20, 1748–1758 (2023).
31. Park, S. et al. Label-free adaptive optics single-molecule localization microscopy

for whole zebrafish. Nat. Commun. 14, 4185 (2023).

32. Cabillic, M. et al. In-depth single molecule localization microscopy using adaptive optics and single objective light-sheet microscopy. Nat. Commun. 16, 8362 (2025).
33. Tavakoli, M. R. et al. Light-microscopy-based connectomic reconstruction of mammalian brain tissue. Nature 642, 398–410 (2025).
34. Michalska, J. M. et al. Imaging brain tissue architecture across millimeter to nanometer scales. Nat. Biotechnol. 42, 1051–1064 (2024)
35. Gao, R. et al. Cortical column and whole-brain imaging with molecular contrast and nanoscale resolution. Science 363, eaau8302 (2019).
36. Grahammer, F., Schell, C. & Huber, T. B. The podocyte slit diaphragm—from a thin grey line to a complex signalling hub. Nat. Rev. Nephrol. 9, 587–598 (2013).
37. Kocylowski, M. K. et al. A slit-diaphragm-associated protein network for dynamic control of renal filtration. Nat. Commun. 13, 6446 (2022).
38. Siegerist, F. et al. Structured illumination microscopy and automatized image processing as a rapid diagnostic tool for podocyte effacement. Sci. Rep. 7, 11473 (2017).
39. Suleiman, H. Y. et al. Injury-induced actin cytoskeleton reorganization in podocytes revealed by super-resolution microscopy. JCI Insight 2, e94137 (2017).
40. Tesch, F. et al. Super-resolved local recruitment of CLDN5 to filtration slits implicates a direct relationship with podocyte foot process effacement. J. Cell. Mol. Med. 25, 7631–7641 (2021).
41. Li, L., Gattass, R. R., Gershgoren, E., Hwang, H. & Fourkas, J. T. Achieving λ/20 resolution by one-color initiation and deactivation of polymerization. Science 324, 910–913 (2009).
42. Wollhofen, R., Katzmann, J., Hrelescu, C., Jacak, J. & Klar, T. A. 120 nm resolution and 55 nm structure size in STED-lithography. Opt. Express 21, 10831–10840 (2013).
43. Xie, F., Song, S., Liang, L., Li, X. & Cao, Y. Sub-100 nm pixel pitch via STED photolithography with a nanoprinting-at-expansion/employments-at-recovery strategy. Opt. Express 31, 2892–2901 (2023).
44. Gould, T. J., Burke, D., Bewersdorf, J. & Booth, M. J. Adaptive optics enables 3D STED microscopy in aberrating specimens. Opt. Express 20, 20998–21009 (2012).
45. Patton, B. R. et al. Three-dimensional STED microscopy of aberrating tissue using dual adaptive optics. Opt. Express 24, 8862–8876 (2016).
46. Antonello, J., Barbotin, A., Chong, E. Z., Rittscher, J. & Booth, M. J. Multi-scale

sensorless adaptive optics: application to stimulated emission depletion microscopy. Opt. Express 28, 16749–16763 (2020).
47. Wang, T. et al. Three-photon imaging of mouse brain structure and function through the intact skull. Nat. Methods 15, 789–792 (2018).
48. Streich, L. et al. High-resolution structural and functional deep brain imaging using adaptive optics three-photon microscopy. Nat. Methods 18, 1253–1258 (2021).
49. Rodríguez, C. et al. An adaptive optics module for deep tissue multiphoton imaging in vivo. Nat. Methods 18, 1259–1264 (2021).
50. Qin, Z. et al. Deep tissue multi-photon imaging using adaptive optics with direct focus sensing and shaping. Nat. Biotechnol. 40, 1663–1671 (2022).
51. Ries, J. SMAP: a modular super-resolution microscopy analysis platform for SMLM data. Nat. Methods 17, 870–872 (2020).
52. Nieuwenhuizen, R. P. J. et al. Measuring image resolution in optical nanoscopy. Nat. Methods 10, 557–562 (2013).

# Methods

## Focal plane extreme adaptive optics

A detailed derivation of the FP-ExAO framework is provided in Supplementary Note 1.

### Multi conjugate AO forward model and linearized state space form

Under the standard scalar, paraxial Fourier optics description, let $E_{ab}$ denote the complex field at the objective back pupil before DM modulation. DM1 is placed at the pupil plane, and DM2 is located at a plane separated from the pupil plane by an axial distance $z$. The modulation and propagation through the two DMs can be written as:

$$E_p = E_{ab} \cdot exp\{i\varphi_1\} \cdot H_z\{exp\{i\varphi_2\}\}, \qquad E_f = \mathcal{F}\{E_p\} \tag{1.1}$$

Here, $E_p$ denotes the complex field at the pupil plane after DM modulation, $E_f$ denotes the complex field at the focal plane, $H_z$ denotes the free space propagation operator over distance $z$, and $\mathcal{F}$ denotes the Fourier transform operator.

Each DM is modeled as a reflective phase modulator, and its surface is approximated by a linear superposition of actuator influence functions:

$$\varphi = 2 \cdot \frac{2\pi}{\lambda} \cdot \phi, \qquad \phi = \sum_{q=1}^{N_{act}} u_q f_q \tag{1.2}$$

where $\varphi$ denotes the phase modulation introduced by the DM, $\phi$ denotes the DM surface height, $u_q$ is the control voltage applied to the $q$-th actuator, $f_q$ is the corresponding influence function, and $N_{act}$ is the total number of actuators.

If the per-iteration control voltage increment $\Delta u_k$ is sufficiently small, the incremental phase term induced by $\Delta u_k$ is linearized by a first order Taylor expansion, yielding the linearized field update:

$$E_{f,k} = E_{f,k-1} + G\Delta u_k + \omega_{f,k} \tag{1.3}$$

where $G$ denotes the system Jacobian matrix and $\omega_{f,k}$ denotes the process noise.

The system Jacobian $G$ is defined as the partial derivative of the complex focal-plane field $E_f$ with respect to the actuator voltages $u$, characterizing how infinitesimal voltage changes map to variations in the complex focal-plane field:

$$G \triangleq \frac{\partial E_f}{\partial u} \tag{1.4}$$

For subsequent estimation and control in the real domain, the complex field is converted into an augmented real-valued representation:

$$x_{f,k} = x_{f,k-1} + G_R \Delta u_k + \omega_{f,k}$$
$$G_R \triangleq \begin{bmatrix} \Re\{G\} \\ \Im\{G\} \end{bmatrix}, \qquad x_{f,k} \triangleq \begin{bmatrix} \Re\{E_{f,k}\} \\ \Im\{E_{f,k}\} \end{bmatrix} \tag{1.5}$$

**Optimal electric field estimators and controllers**

At the $k$-th AO iteration, to estimate the focal-plane field and correct aberrations, $M$ pairs of symmetric probe commands $\pm u_k^p$ are applied on the pupil plane DM. According to Eq. 1.3, the focal-plane complex fields under the $m$-th positive and negative probe are:

$$E_{f,k}^{m+} = E_{f,k} + G u_k^{p,m},\ E_{f,k}^{m-} = E_{f,k} - G u_k^{p,m}, \qquad \forall\, m = 1,2,\dots,M \tag{1.6}$$

where $E_{f,k}^{m+}$ denotes the complex focal-plane field under the $m$-th positive probe and $E_{f,k}^{m-}$ denotes the corresponding field under the negative probe.

The focal-plane intensity measurements are:

$$I_{f,k}^{m+} = E_{f,k}^{m+^*} E_{f,k}^{m+} + n_{f,k}^{+}, \qquad I_{f,k}^{m-} = E_{f,k}^{m-^*} E_{f,k}^{m-} + n_{f,k}^{-} \tag{1.7}$$

For each probe pair, the differential intensity observation is computed and stacked as:

$$\begin{bmatrix} \Delta I_k^1 \\ \vdots \\ \Delta I_k^M \end{bmatrix} = \begin{bmatrix} I_{f,k}^{1+} - I_{f,k}^{1-} \\ \vdots \\ I_{f,k}^{M+} - I_{f,k}^{M-} \end{bmatrix} = 4\Re\left\{ \begin{bmatrix} (G u_k^{p,1})^* \circ E_{f,k} \\ \vdots \\ (G u_k^{p,M})^* \circ E_{f,k} \end{bmatrix} \right\} + n_{f,k} \tag{1.8}$$

which can be simplified as:

$$\delta_k = A x_{f,k} + n_{f,k} \tag{1.9}$$

where:

$$\delta_k = \begin{bmatrix} \Delta I_k^1 \\ \vdots \\ \Delta I_k^M \end{bmatrix}, \qquad A = 4 \begin{bmatrix} diag\left(\Re\{G u_k^{p,1}\}\right) & diag\left(\Im\{G u_k^{p,1}\}\right) \\ & \vdots \\ diag\left(\Re\{G u_k^{p,M}\}\right) & diag\left(\Im\{G u_k^{p,M}\}\right) \end{bmatrix} \tag{1.10}$$

Under this linear observation model, focal-plane field estimation is formulated as a least squares problem, whose closed-form solution is:

$$\hat{x}_{f,k} = (A^T A)^{-1} A^T \delta_k \tag{1.11}$$

To drive the point source PSF toward an ideal shape, the control increment is computed by suppressing energy outside the PSF core, thereby promoting energy concentration into the center. Specifically, the binary mask $M$ is defined according to the desired focal-field distribution for each imaging modality. For SMLM, where the objective is to concentrate the PSF energy into the central region, pixels within a radius $r_0$ are assigned $M = 0$, whereas pixels outside $r_0$ are assigned $M = 1$, such that minimizing the masked energy suppresses the peripheral field while preserving the central lobe. For STED, the mask is instead designed according to the desired doughnut shaped depletion focus: pixels within the target annular region, $r_{in} \leq r \leq r_{out}$, are assigned $M = 0$, while pixels outside this annulus are assigned $M = 1$. Minimizing the corresponding masked energy therefore suppresses both the central and outer region energy, effectively concentrating the optical energy into the desired doughnut ring while maintaining a low intensity center. The control increment is obtained by minimizing the following loss:

$$\Delta u_k = \underset{\Delta u}{\operatorname{argmin}} \left\| \sqrt{W} (\hat{x}_{f,k} + G_R \Delta u) \right\|_2^2 + \sqrt{\alpha} \|\Delta u\|_2^2 \tag{1.12}$$

where the first term penalizes energy outside the ROI and the second term is a Tikhonov regularizer that constrains the control amplitude.

For the regularized least squares problem defined in Eq. 1.12, the closed-form solution of the control increment is:

$$\Delta u_k = -( {G_R}^T W \, G_R + \alpha I)^{-1} \, {G_R}^T W \hat{x}_{f,k} \tag{1.13}$$

For FP-ExAO (cascaded DMs) correction of field dependent aberrations, the field of view is partitioned into $N \times N$ isoplanatic patches, each associated with its own Jacobian. Within each patch, PSFs are detected, cropped, and averaged to form a representative PSF. A joint loss is defined by minimizing the out of core energy over all patches, enabling collaborative optimal correction of spatially varying aberrations. The control increment is computed as:

$$\Delta u_k = \underset{\Delta u}{\operatorname{argmin}} \sum_{i=1}^{N \times N} \left\| \sqrt{W} (x_{f,k}^i + G_R^i \Delta u) \right\|_2^2 + \sqrt{\alpha} \|\Delta u\|_2^2 \tag{1.14}$$

For the least squares problem defined in Eq. 1.14, the closed-form solution of the control increment is:

$$\Delta u_k = -\left( \sum_{i=1}^{N \times N} {G_R^i}^T W G_R^i + \alpha I \right)^{-1} \left( \sum_{i=1}^{N \times N} {G_R^i}^T W \hat{x}_{f,k}^i \right) \tag{1.15}$$

**Calibration of the system Jacobian matrix**

The real-valued Jacobian $G_R$ enters directly into focal-plane complex field estimation and the computation of control voltages; any bias in $G_R$ can therefore substantially degrade convergence speed and closed loop stability. In experiments, an initial Jacobian $G_R^{(0)}$ is provided by an optical simulation model and is subsequently refined in situ using known DM voltage perturbations and the corresponding differential intensity measurements. From Eq. 1.9, the difference between differential observations at two consecutive iterations is:

$$\Delta\delta_k = \delta_k - \delta_{k-1} = AG_R\Delta u_k = C\Delta u_k \tag{1.16}$$

Calibration data are collected over multiple voltage perturbations by stacking $\Delta u_k$ and the corresponding $\Delta\delta_k$, from which the regularized least squares estimate $\hat{C}$ is obtained. Using the simulation derived Jacobian $G_R^{(0)}$ as initialization, the in situ calibrated Jacobian is then obtained by solving:

$$\hat{G}_R = \underset{G_R}{\operatorname{argmin}} \left\|\hat{C} - AG_R\right\|_2^2 + \mu\left\|G_R - G_R^{(0)}\right\|_2^2 \tag{1.17}$$

using a numerical optimizer such as L-BFGS. A detailed derivation and implementation of the calibration procedure are provided in Supplementary Note 1.

**Optimal probe**

The probe type directly determines whether focal-plane field estimation is information rich and numerically stable. For the linear observation model $\delta_k = Ax_{f,k} + n_{f,k}$, probe design amounts to constructing $u_k^p$ such that the probe-dependent sensing matrix $A$ is maximally informative while remaining well conditioned. Assuming Gaussian noise, the Fisher information matrix is $F(x_{f,k}) = A^T R^{-1} A$, with $R = \sigma^2 I$, and the probes are selected according to the D-optimal criterion that maximizes $\log\det \mathrm{F}(\mathrm{x}_{\mathrm{f,k}})$. For two probe pairs, the pixel wise information determinant is:

$$detF_j(x_{f,k}) = \frac{16^2}{\sigma^4}\left|p_{k,j}^1\right|^2\left|p_{k,j}^2\right|^2 \sin^2(\phi_{k,j}^2 - \phi_{k,j}^1) \tag{1.18}$$

Maximizing determinant requires approximately equal probe amplitudes $\left|p_{k,j}^1\right| = \left|p_{k,j}^2\right|$ and a phase separation $\phi_{k,j}^2 - \phi_{k,j}^1 = \pm\frac{\pi}{2}$ at each pixel, yielding quadrature probe responses. Because the probes are applied at the pupil plane, an approximately uniform and non-vanishing perturbation within the focal-plane ROI additionally requires sinc-like probe commands. In this work, four pairs of orthogonal sinc-like probes were constructed using the Hilbert transform. The complete fisher information analysis and the explicit construction of the four probe commands are provided in Supplementary Note 1.

**Optical setups of FP-ExAO SMLM microscope**

As shown in Extended Fig. 7, FP-ExAO SMLM microscope was constructed by integrating a homogenized wide field excitation path, a multi conjugate adaptive optics relay in the detection arm, and a near infrared focus lock module. All subsystems shared the same imaging objective and a common mechanical reference to ensure long term stability during SMLM acquisition.

Two excitation lasers at 405 nm (OBIS405 LX, 140 mW, Coherent) and 647 nm (2RU-VFL-P-2000-647-B1R, 2 W, MPB Lasers) were combined using mirrors and long-pass dichroic mirrors DiM1-3 (DM10-425/505/605LP, Lbtek). The combined beam was coupled into a square core multimode fiber (WFRCT200x200/230x230/440/620/1100N, NA 0.22, CeramOptec) mounted on a fiber agitator to reduce spatial coherence and suppress speckle. The fiber output was collimated by OL1 (PLN10X, Olympus) and relayed to the microscope body by lenses L2-L5 (f = 150 mm, f = 30 mm, f = 300 mm, f = 400 mm). An excitation cleanup filter (ZET405/488/561/640xv2, Chroma) was inserted after the fiber relay to suppress fiber induced background fluorescence. The excitation beam was then reflected by a multiband dichroic mirror (ZET405/488/561/640mv2, Chroma) into a silicone oil immersion objective (UPLSAPO100XS, NA 1.35, Olympus), providing homogeneous wide field illumination over a 75 μm × 75 μm field of view at the sample plane (Supplementary Fig. 2).

Fluorescence emission was collected by the same objective and transmitted through the tube lens and L9 (f = 180 mm). The emission path contained a multi conjugate AO relay consisting of two deformable mirrors. The first deformable mirror, DM1 (DM140, Multi-3.5, Boston Micromachines Corp), was placed at the pupil plane and mainly corrected spatially invariant aberrations. The second deformable mirror, DM2 (DM952, Kilo-C-3.5, Boston Micromachines Corp), was placed downstream with an inter DM separation of 146 mm, as determined by optical simulation under the constraints of actuator aperture, field size, and relay geometry. This second mirror provided an additional correction degree of freedom for compensating spatially varying aberrations. After AO modulation, the fluorescence was relayed to an sCMOS camera (ORCA-Flash4.0 V3, C13440-20CU, Hamamatsu) through the relay lenses L10-L12 (f = 400 mm, f = 300 mm, f = 60 mm) and an emission filter (ZET405/488/561/647m, Chroma). The resulting total magnification was 53.3×, corresponding to an effective pixel size of 122 nm at the sample plane.

Axial stabilization was implemented using an 808 nm near infrared focus lock beam (TEM-F-808nm-20mW, CNIlaser) delivered through a single mode fiber. The beam was collimated by L7 (f = 25 mm), split into two laterally separated beams, passed through L6 (f = 400 mm), and then coupled into the objective through a polarizing beam splitter, a quarter wave plate (QWP25-808A-M, Lbtek), and an infrared dichroic mirror DiM4 (ZT775sp-2p, Chroma). After reflection at the sample glass interface and double passage through the quarter wave plate, the returning beams were directed by the polarizing beam splitter to a lock camera (MER2-160-227U3M, Daheng Imaging). The relative displacement of the two reflected spots encoded axial sample drift and was used as feedback to drive the XYZ stage (MS-2000 XYZ, ASI) for real time focus stabilization (Supplementary Fig. 1).

**Optical setups of STED nanolithography microscope**

As shown in Extended Fig. 8, the STED direct laser writing system was constructed by combining a femtosecond excitation path, a vortex depletion and wavefront control path, a dual axis beam scanning module, and an in-situ transmission monitoring module. The excitation and depletion beams were combined before the scanning module and delivered through the same high-NA objective, ensuring common lateral scan trajectories. The target writing depth was set by the motorized z stage and was defined as the axial distance from the coverslip–photoresist interface to the excitation focal plane.

A 780 nm femtosecond laser (Insight X3+, 120 fs, 80 MHz) was used as the excitation source. The excitation beam first passed through an optical attenuator and shutter S1, and was then spatially filtered and expanded using L1 (f = 40 mm), a 20 μm pinhole, and L2 (f = 200 mm). Its polarization was adjusted using a half wave plate (HWP) and a quarter wave plate (QWP). The beam was subsequently directed by mirrors M1 and M2 to a dichroic mirror DM (775sp, Chroma), where it was combined with the depletion beam.

A 532 nm continuous wave laser (LWGL532-500mW, Laserwave) was used as the depletion source. The depletion beam passed through shutter S2 and was directed by a prism onto a reflective spatial light modulator (SLM; PLUTO-VIS-016, Holoeye), which simultaneously imposed the first-order vortex phase required for generation of the doughnut-shaped depletion beam and the FP-ExAO correction phase. Thus, structured-field generation and aberration correction were implemented using the same phase-only device. The modulated beam was relayed through L11 (f = 300 mm), mirrors M6 and M5, L10 (f = 40 mm), a half wave plate (ZWP20H-532Q, JCOPTIX), a quarter wave plate (QWP20-532B, Lbtek), L9 (f = 250 mm), mirror M4, and L8 (f = 250 mm). The half wave plate and quarter wave plate were used to generate left-handed circular polarization required to maintain a rotationally symmetric central null under high-NA focusing.

The excitation and depletion beams were combined at dichroic mirror and subsequently scanned laterally by two single axis galvanometers. The scanning relay consisted of two scan lens and tube lens pairs, with L3 and L5 serving as scan lenses (CLS-SL, Thorlabs) and L4 and L6 as tube lenses (TTL200MP, Thorlabs). Mirror M3 folded the optical path between the two galvanometers. After scanning, the co-aligned beams were directed by BS1 into the objective lens (Olympus, LUMFLN, 60×, NA = 1.1, WD = 1.5 mm) and focused into the photoresist. Lateral patterning was performed with the galvanometers, whereas the three-axis motorized stage (MAX311D, Thorlabs) controlled the sample position and writing depth.

For in situ transmission monitoring, the sample was illuminated from below using a 650 nm LED. The transmitted light was collected by the same objective, separated from the writing beams by BS1 and BS2, and imaged through L7 (f = 200 mm) onto a camera (MC050MG-SY, XIMEA). This transmission monitoring path enabled real time observation of the sample during direct laser writing.

**Instrument control and AO driven SMLM data acquisition**

AO correction and focus locking were performed on a workstation using Python based instrument control. During AO correction, the deformable mirrors, lasers, scientific camera, focus lock camera, and motorized stage were coordinated to execute probe image acquisition, focal plane field estimation, DM command update, and real time axial stabilization. After AO correction was completed and focus lock was engaged, raw SMLM movies were acquired from the Hamamatsu scientific camera using the Hamamatsu DCAM python SDK and saved for subsequent single molecule localization analysis.

For SMLM acquisition, the 647 nm excitation power density was first increased to 10–25 kW/cm$^2$ to pre-bleach the sample and drive it into a sparse blinking regime, and was then reduced 2–6 kW/cm$^2$ for AO estimation and correction. Under this condition, a frame sequence was recorded for one AO loop processing (2–5 loops were used for AO): each frame was bias corrected, a temporal median background was estimated and subtracted, and PSF candidates were detected and cropped into subimages. The resulting PSF subimage sequence served as the input to FP-ExAO (pupil DM)/FP-ExAO (cascaded DMs) to form pairwise differential measurements and update the control voltages (For FP-ExAO (pupil DM), the conjugate DM was held at the flatmap, whereas for FP-ExAO (cascaded DMs) both DMs were updated according to their respective correction commands). After AO convergence, the optimized DM commands were recorded and used for interleaved SMLM acquisition. For FP-ExAO (cascaded DMs) experiments, the DM commands were switched among FP-ExAO (cascaded DMs), FP-ExAO (pupil DM), and No AO states, with 2,000 frames acquired for each state before switching to the next condition. For FP-ExAO (pupil DM) experiments, the DM commands were switched between FP-ExAO (pupil DM) and No AO states, also with 2000 frames acquired per state. The focus lock module remained active throughout the acquisition. Data acquisition and computational processing were performed on the same workstation equipped with an Intel Core i9-10980XE CPU (18 cores, 3.00 GHz) and an NVIDIA GeForce RTX 4060 GPU with 8 GB of dedicated memory.

**3D SMLM localization and reconstruction**

For 3D SMLM imaging, RMS = 60 nm astigmatism was introduced by applying an astigmatic mode to the deformable mirror. Under the same astigmatic configuration, fluorescent beads immobilized on a coverslip were imaged while performing an axial z scan to acquire an experimental 3D PSF calibration stack. The resulting bead z stack was converted in SMAP into a cspline PSF model. Raw SMLM movies acquired with the same optical configuration were then localized in SMAP using the calibrated spline PSF model, and the resulting localization lists were rendered in SMAP to generate the final 3D super resolution reconstructions[51]. Fourier ring correlation (FRC) was used to estimate reconstructed image resolution, using the conventional 1/7 threshold criterion[52]. Detailed localization metrics and FRC resolution values for the reconstructed SMLM datasets are summarized in Supplementary Tables 1 and 2.

## Sample preparation

### Imaging buffer and sample mounting for single molecule localization microscopy

All imaging buffers were prepared immediately before use. The conventional STORM imaging buffer consisted of standard imaging buffer (STIB) supplemented with 1% (v/v) β-mercaptoethanol (β-ME) and 1% (v/v) glucose oxidase-catalase oxygen-scavenging system (GLOX). STIB contained 100 mM Tris-HCl (pH 8.0), 20 mM NaCl, and 10% (w/v) glucose, and the GLOX stock contained 60 mg/ml glucose oxidase, 6 mg/ml catalase, and 40% (v/v) glycerol. The refractive index matched STORM imaging buffer consisted of 38.1% (w/v) sucrose in 1× STIB, supplemented with 1% (v/v) β-ME and 1% (v/v) GLOX. The refractive index of the final buffer was measured at 23 °C as (n=1.4062) using a DR6000-T digital refractometer (A. KRÜSS Optronic GmbH, Hamburg, Germany; 589 nm light source and Peltier temperature control), closely matching the refractive index of the Olympus silicone immersion oil used for imaging (n=1.406).

For imaging, labelled tissue sections were transferred to a glass bottomed dish and gently flattened (adherent cell samples were cultured directly on glass bottomed dishes and therefore did not require transfer). A 50–80 µL aliquot of freshly prepared imaging buffer was applied to each sample, and the sample was overlaid with a coverslip. The edges of the coverslip were sealed with nail polish to minimize oxygen exchange, and SMLM acquisition was initiated immediately after mounting.

### Fixation and immunolabeling of TOM20 in COS-7 cells

COS-7 cells were cultured on pretreated glass bottom dishes until reaching approximately 60% confluence. The cells were fixed with 3% paraformaldehyde (PFA) and 0.1% glutaraldehyde (GA) in PBS, permeabilized, treated with freshly prepared $NaBH_4$ to reduce aldehyde induced fluorescence background, and blocked with BSA. TOM20 was labeled with rabbit anti-TOM20 primary antibody (Abcam, Cat. ab78547; 1:1000) overnight at 4 °C, followed by Alexa Fluor 647-conjugated goat anti-rabbit IgG (Thermo Fisher Scientific, Cat. A-21245; 1:2000) for 2 h at room temperature.

### Mouse tissue collection, fixation, and sectioning

Mice were obtained from Beijing Vigorstar Biotechnology Co., Ltd. The animals were housed under a 12-h light/dark cycle and had random access to food and water. The temperature was maintained at 18–23 °C, and humidity was at 40–60%. The animal experiments were conducted according to institutional guidelines of Peking University.

Mice used for fixed tissue imaging were deeply anesthetized and transcardially perfused with PBS\heparin solution (10–15 U heparin per milliliter) followed by 4% PFA. The brains and kidneys were dissected, post fixed in 4% PFA at 4 °C overnight, and thoroughly washed with PBS for 3 times. Brain tissues were sectioned at a thickness of 200 μm, whereas kidney tissues were sectioned at thicknesses of 20 or 60 μm using a vibratome (VT1200S, Leica).

**Immunolabeling of GFAP and GFP in mouse brain sections**

To label GFAP-positive astrocytic structures in brain tissue, brain sections were obtained from 9-week-old C57BL/6J mice. To reduce fixation associated background fluorescence, the sections were first treated with freshly prepared 0.1 M $NH_4Cl$ for 10 min. The sections were subsequently permeabilized with 0.4% Triton X-100 and blocked in PBS containing 5% BSA, 10% normal goat serum and 0.04% Triton X-100 before being incubated overnight at 4 °C with rabbit anti-GFAP antibody (Novus Biologicals/Bio-Techne, Cat. no. NB300-141; 1:200). After five washes with PBS, the sections were incubated for 2 h at room temperature with Alexa Fluor 647-conjugated goat anti-rabbit IgG (Thermo Fisher Scientific, Cat. no. A-21245; 1:200), followed by five additional PBS washes.

To label GFP expressing neuronal structures in brain tissue, brain sections were obtained from 9-week-old Thy1-GFP mice. To reduce background fluorescence, the sections were first treated with freshly prepared 0.1% (w/v) $NaBH_4$ for 5 min. The sections were subsequently permeabilized with 0.4% Triton X-100 and blocked in PBS containing 5% BSA and 0.01% Triton X-100 before being incubated overnight at 4 °C with Alexa Fluor 647-conjugated rabbit anti-GFP antibody (Invitrogen/Thermo Fisher Scientific, Cat. no. A-31852; 1:200). The sections were then washed five times with PBS.

**Immunolabeling of Nephrin in mouse kidney sections**

Kidney sections were obtained from 8-week-old C57BL/6J mice. To reduce fixation associated background fluorescence, the sections were first treated with freshly prepared 0.1 M $NH_4Cl$ for 10 min. The sections were subsequently permeabilized with 0.4% Triton X-100 and blocked in PBS containing 5% BSA, 10% normal donkey serum and 0.04% Triton X-100 before being incubated overnight at 4 °C with sheep anti-human Nephrin antibody (R&D Systems/Bio-Techne, Cat. no. AF4269-SP; 50 μg $ml^{-1}$). After washing with PBS, the kidney sections were incubated for 1 h at room temperature with Alexa Fluor 647-conjugated donkey anti-sheep IgG (Thermo Fisher Scientific, Cat. no. A-21448; 1:200).

**Establishment of the adriamycin induced FSGS mouse model**

Male 12-week-old C57BL/6J mice were divided into control, low dose, and high dose groups. Adriamycin induced FSGS was established by a single tail vein injection of doxorubicin hydrochloride at 10 mg/kg or 20 mg/kg, whereas control mice received an equivalent volume of saline. The mice were maintained for 28 days after injection and were then euthanized for kidney collection. Kidney tissues were fixed and sectioned as described above for subsequent histological and SMLM analyses.

**Skeletonization and quantitative analysis of nephrin-positive slit diaphragm networks**

Reconstructed SMLM probability density images were cropped into 2000 × 2000 pixels regions with a 10 nm pixel size, corresponding to a 20 μm × 20 μm field of view. Images from the Healthy FP-ExAO, Healthy NoAO, mild FSGS FP-ExAO and severe FSGS FP-ExAO groups were processed with the same parameter set.

For skeleton extraction, each cropped image was first smoothed with a Gaussian filter. A low spatial frequency background was estimated and subtracted from the smoothed image, and the corrected image was normalized by the background noise level estimated from low intensity pixels. Hysteresis thresholding was then used to generate a binary nephrin-positive mask. Small connected objects were removed, and compact or non filamentous objects were excluded according to their skeleton length and mean width. No morphological closing or gap bridging operation was applied, so discontinuities present in pathological samples were not artificially connected.

The retained binary line mask was reduced to a one pixel wide skeleton. For visualization only, the skeleton was dilated and displayed in red. All quantitative measurements were performed on the original one pixel wide skeleton. The skeleton was represented as an eight connected pixel graph, and the total skeleton length was obtained by summing the physical lengths of horizontal, vertical and diagonal graph edges while avoiding double counting of diagonal shortcuts.

Three metrics were extracted from each skeleton. Slit diaphragm density (SD) was calculated as the skeleton length per analyzed area:

$$SD = \frac{L_{corr}}{A_{ROI}} \tag{1.19}$$

where $L_{corr}$ is the total skeleton length after endpoint trimming and $A_{ROI}$ is the analyzed image area. Endpoint trimming was used to reduce overestimation from short terminal branches in fragmented skeletons.

The nephrin continuity ratio (NCR) was defined as the fraction of total skeleton length contributed by long connected components:

$$NCR = \frac{\sum_i L_i \mathbb{I}(L_i \geq 1.0\mu m)}{\sum_i L_i} \tag{1.20}$$

where $L_i$ is the length of connected component $i$, and $\mathbb{I}(\cdot)$ is the indicator function. Higher NCR values indicate that a larger fraction of the nephrin-positive network is contained in long, continuous structures.

Gap density was estimated from internal skeleton endpoints:

$$D_{gap} = \frac{N_{end}}{2L_{dot}} \times 10 \tag{1.21}$$

where $N_{end}$ is the number of internal skeleton endpoints and $L_{dot}$ is the total skeleton length before endpoint trimming. The value reports the estimated number of gaps per 10 μm of nephrin-positive skeleton. Higher gap density indicates greater network fragmentation. All segmentation thresholds, component filtering rules and quantitative parameters were fixed before group comparison and kept identical across all groups.

**FP-ExAO for STED nanolithography**

STED nanolithography was performed using a photoinhibitable resist composed of 7-diethylamino-3-thenoylcoumarin (DETC) as the photoinitiator and pentaerythritol tetraacrylate (PETTA) as the monomer. DETC and PETTA were mixed at a mass ratio of 0.0025:1. The prepared resist was stored in the dark at 4 °C before use. Samples were prepared in a glass slide–photoresist–coverslip configuration. Glass slides and coverslips were cleaned sequentially with acetone and ethanol and dried before assembly. A hollow spacer was attached to the glass slide to define the resist chamber, which was filled with photoresist and sealed with a coverslip. During STED direct laser writing, two photon absorption of 780 nm femtosecond pulse laser drives the photoinitiator into the excited state and initiates polymerization, whereas the spatially structured 532nm depletion beam returns excited molecules to the ground state before they contribute to polymerization initiation by stimulated emission. Because the depletion field contains a central intensity null surrounded by a high-intensity annulus, polymerization is retained near the centre while being inhibited in the surrounding region. The effective writing voxel is therefore determined jointly by the excitation dose, the central residual depletion intensity and, critically, the annular depletion intensity present at the excitation focal plane.

For FP-ExAO correction, adaptive optics control was applied only to the depletion beam because preservation of the central intensity minimum and annular symmetry of the vortex depletion focus directly determines the lateral confinement of the polymerization region. The excitation path was left unchanged. The spatial light modulator (SLM) in the depletion arm simultaneously generated the vortex phase required for the STED depletion focus and applied the FP-ExAO correction phase. The system response was first calibrated using the procedure described above for FP-ExAO. Pairwise probe perturbations were subsequently applied through the SLM, and the corresponding focal-plane intensity measurements were used to estimate the complex depletion field. The estimated field and calibrated system response were then used to compute the corrective SLM phase using the same FP-ExAO estimation and closed-form control framework described above. Because nanolithography was performed within a single writing region rather than over a spatially distributed imaging field, only a single region FP-ExAO correction was required and no field partitioning or multi region optimization was performed. After convergence, the optimized correction phase was combined with the vortex phase and retained on the SLM throughout fabrication. For NoAO measurements, the same optical configuration and vortex phase were used without the FP-ExAO correction phase.

After AO correction, the excitation and depletion foci were spatially overlapped inside the photoresist and the focal plane was positioned at the prescribed writing depth. Structures were then written at a depth of 160 μm using an excitation power of 13.4 mW, a depletion power of 7.26 mW and a scan speed of 60 μm $s^{-1}$. Optical powers were measured at sample plane and were held constant across the FP-ExAO, metric-based AO and NoAO conditions. The complex patterns were generated from identical trajectories under all correction conditions. After writing, the samples were developed by sequential immersion in isopropanol for 15 min, acetone for 15 min and ethanol for 15 min, followed by air drying to remove unpolymerized and residual photoresist.

The developed structures were characterized by scanning electron microscopy (SEM)

using a Helios NanoLab 600i scanning electron microscope at an accelerating voltage of 2 kV. Linewidths and inter line gaps were measured from spatially calibrated SEM images along directions perpendicular to the written structures. For line grating and complex pattern comparisons, intensity profiles were extracted along identical representative trajectories from the AO and NoAO SEM images and normalized for visualization. The same image processing and measurement procedures were applied to both conditions.

## Acknowledgments

We thank Prof. Kun Song from Peking University and the core facility of the National Biomedical Imaging Center, Peking University, for their support in tissue collection and sectioning. We also thank the core facility of the School of Physics, Peking University, for providing scanning electron microscopy (SEM) support.

## Funding

This work was supported by grants from the National Natural Science Foundation of China (32450631, 62371007, 12621002), the National Key Research and Development Program of China (2024YFC3406400, 2022YFC3401100), and the Natural Science Foundation of Beijing Municipality (Z240010).

## Author contributions

J.L. and H.S. conceived the original idea. J.L. and H.S. designed the experiments. K.S., Y.W., S.Z., Y.S., and H.S. supervised the work. J.L. developed the FP-ExAO algorithm. J.L. built the SMLM FP-ExAO system. Z.L. built the STED FP-ExAO system. J.L. performed the simulations and fluorescent bead experiments. J.L. performed the SMLM experiments. K.Y. and Z.L. performed the STED experiments. J.N. and X.G. performed the staining of all biological samples. R.Z. and J.R. collected the mouse tissue samples and established the disease models. Z.L. prepared the photoresist and performed scanning electron microscopy (SEM) characterization. J.L. analyzed all experimental data. All authors contributed to writing and revising the manuscript.

## Competing interests

The authors declare no competing interests.

## Data and code availability

Data and code are available from the corresponding authors upon reasonable request.

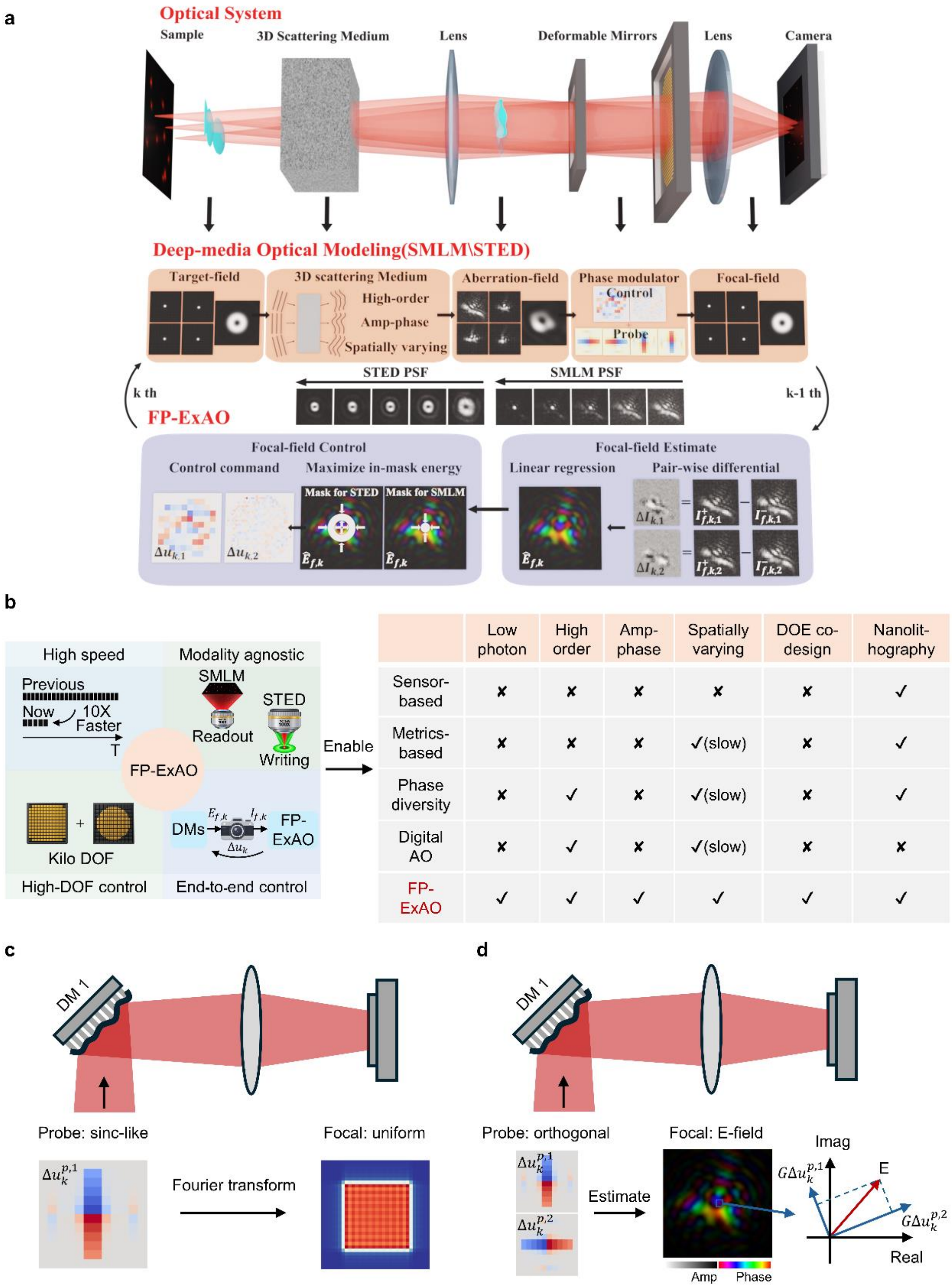

| | Low photon | High order | Amp-phase | Spatially varying | DOE co-design | Nanolit-hography |
|---|---|---|---|---|---|---|
| Sensor-based | ✘ | ✘ | ✘ | ✘ | ✘ | ✓ |
| Metrics-based | ✘ | ✘ | ✘ | ✓(slow) | ✘ | ✓ |
| Phase diversity | ✘ | ✓ | ✘ | ✓(slow) | ✘ | ✓ |
| Digital AO | ✘ | ✓ | ✘ | ✓(slow) | ✘ | ✘ |
| FP-ExAO | ✓ | ✓ | ✓ | ✓ | ✓ | ✓ |

**Figure 1. Concept, workflow and capability comparison of FP-ExAO. a,** Schematic illustration of the multi conjugate adaptive optics system and the FP-ExAO workflow. Probe commands are applied to the deformable mirrors, and pairwise differential focal-plane intensity images are used for optimal field estimation. The estimated complex field PSF and calibrated Jacobian matrix are then used to compute the control command. **b,** Schematic overview of the key capabilities of FP-ExAO, including high degree of freedom (DOF) control, high speed correction, modality agnostic operation and end-to-end closed loop control. These capabilities enable correction of high-order, amplitude–phase coupled and spatially varying aberrations, robust operation under low-photon conditions, nanolithography, and co-design with diffractive optical elements (DOEs). The table compares these capabilities with sensor-based AO, metrics-based AO, phase diversity and digital AO. **c,** Schematic of sinc-like probe design. A sinc-like pupil plane probe is transformed into an approximately uniform focal-plane perturbation within the region of interest. **d,** Schematic of orthogonal optimal probes for focal-plane complex field estimation. The probe pairs measure projections of the focal-plane electric field along different directions in the complex plane. Two orthogonal probe responses provide maximally independent measurements, enabling optimal estimation of the local complex field in both amplitude and phase.

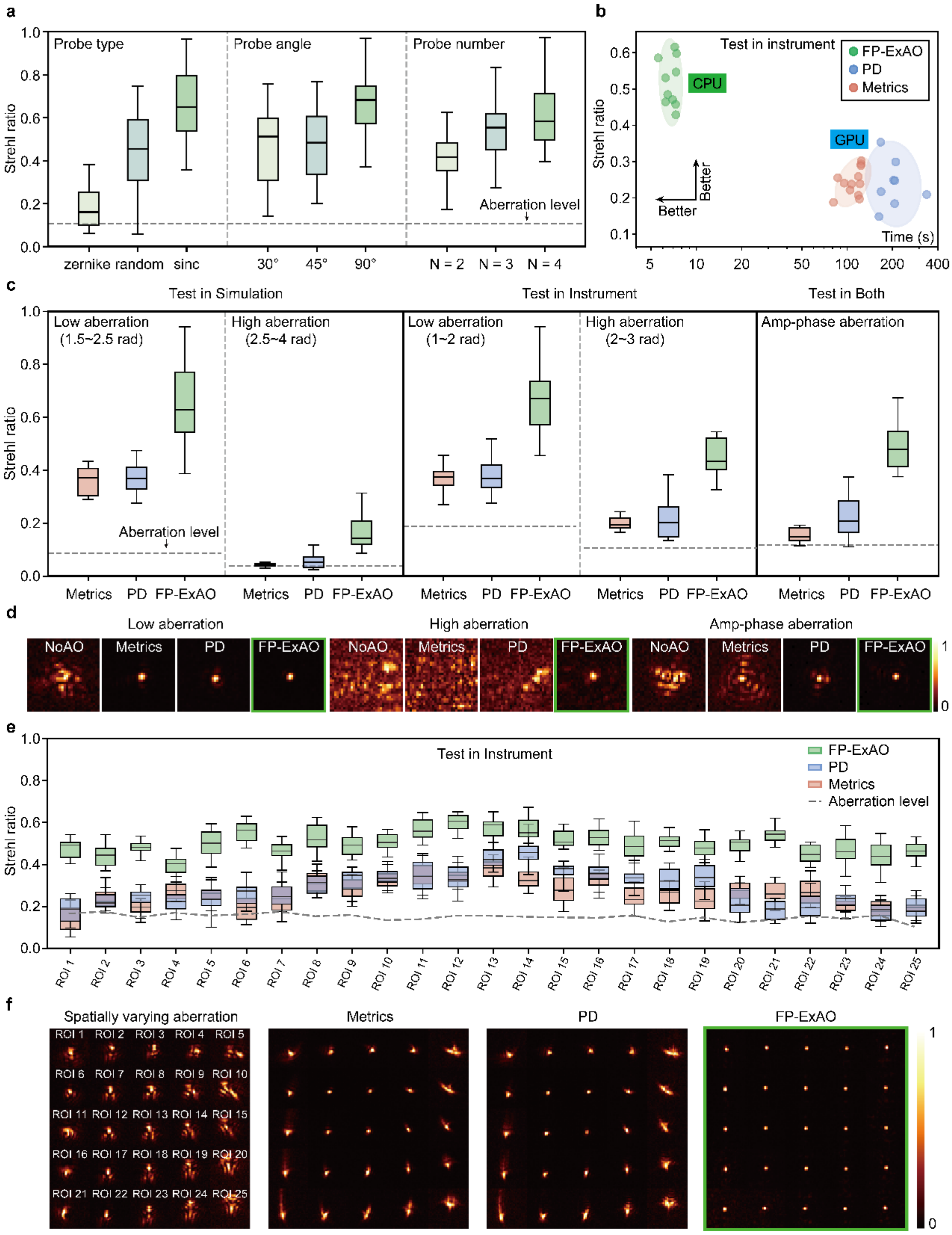

**Figure 2. FP-ExAO probe optimization and correction of randomly introduced spatially varying aberrations. a,** Box plots of Strehl ratios after randomly introduced aberrations were corrected by FP-ExAO using different probe settings. The left group compares probe types, including Zernike, random and sinc probes; the middle group compares probe angles of 30°, 45° and 90°; and the right group compares probe numbers N = 2, 3 and 4. A Strehl ratio of 1 corresponds to the ideal aberration free focus, and the dashed horizontal line indicates the aberrated level before correction. These results identify orthogonal, sinc-like probes as optimal and show that increasing the probe number improves correction robustness. **b,** Scatter plot comparing correction time and corrected Strehl ratio after random aberrations were introduced and then corrected by FP-ExAO, phase diversity AO (PD) or metrics-based AO in instrument tests. The time axis is shown on a logarithmic scale. Each point represents one random aberration correction test. Green, blue and red points correspond to FP-ExAO, PD and Metrics, respectively; CPU and GPU labels indicate the computation devices used in the tests. CPU-based FP-ExAO is more than tenfold faster than the GPU-based conventional methods while achieving higher Strehl ratios. **c,** Box plots comparing the correction of randomly introduced low- and high- level aberrations by Metrics, PD and FP-ExAO in simulation and instrument tests, together with amplitude-phase aberrations tested in both simulation and instrument settings. **d,** Representative normalized PSF images showing the aberrated state and the corrected states after applying Metrics, PD or FP-ExAO under low aberration, high aberration and amplitude-phase aberration conditions. The intensity colour scale is normalized from 0 to 1. All methods effectively correct weak aberrations, whereas FP-ExAO provides superior correction for strong and coupled amplitude–phase aberrations in both simulation and experiment. **e,** Box plots of Strehl ratios measured in 5 × 5 field regions after spatially varying aberrations were introduced and corrected. The horizontal axis lists ROI 1-ROI 25, and the three colours correspond to Metrics, PD and FP-ExAO. The dashed curve indicates the aberrated level before correction. **f,** Representative normalized PSF arrays before and after correction of spatially varying aberrations by Metrics, PD and FP-ExAO. Within each condition, the 25 PSFs are arranged as a 5 × 5 array corresponding to the spatial layout of ROI 1-ROI 25. The four blocks show the introduced spatially varying aberration, Metrics, PD and FP-ExAO, respectively. The intensity colour scale is normalized from 0 to 1. FP-ExAO provides uniform correction across the field, whereas metrics-based AO and PD primarily correct the central region and deteriorate towards the field edges.

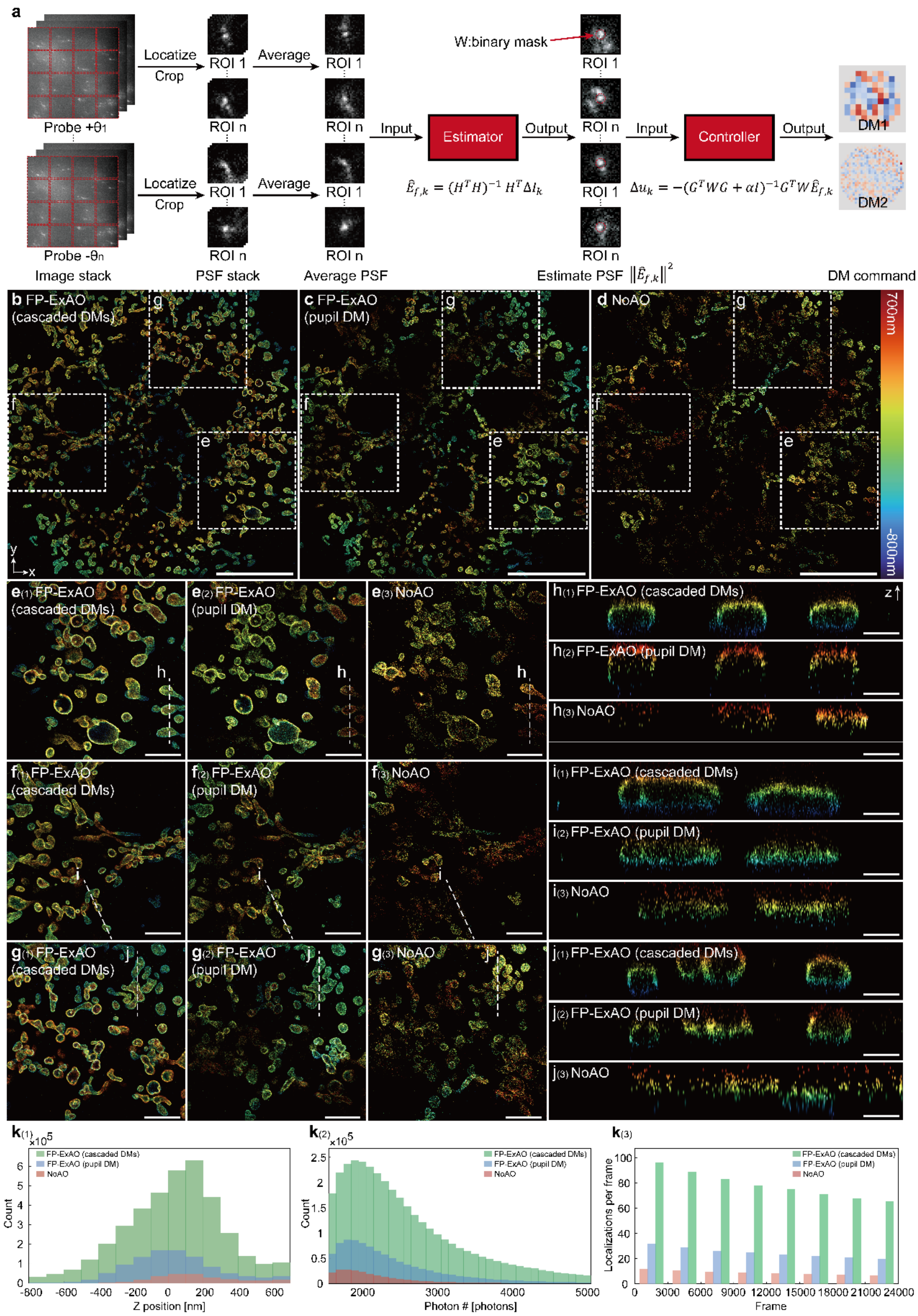

**Figure 3. 3D SMLM imaging of TOM20-labelled mitochondria in COS-7 cells under adaptive optics correction. a,** FP-ExAO workflow for SMLM correction. Probe image stacks acquired under positive and negative probe commands are partitioned into field regions. Blinking PSFs in each region are localized, cropped and averaged to generate representative PSFs. The estimator recovers a focal-plane complex field PSF for each region, and the controller uses the estimated fields, binary weighting mask and calibrated response model to compute commands for the two deformable mirrors. **b-d,** Full field 3D SMLM reconstructions of TOM20-labelled outer mitochondrial membrane structures in COS-7 cells under FP-ExAO (cascaded DMs) (b), FP-ExAO (pupil DM) (c) and NoAO (d) conditions. Aberration was introduced by the deformable mirror. Dashed boxes indicate the regions shown in e-g, and dashed lines indicate the positions of the XZ cross sections shown in h-j. **e-g,** Magnified views of the boxed regions in b-d, comparing TOM20-labelled mitochondrial structures under FP-ExAO (cascaded DMs), FP-ExAO (pupil DM) and NoAO conditions. Subpanels 1, 2 and 3 correspond to FP-ExAO (cascaded DMs), FP-ExAO (pupil DM) and NoAO, respectively. Aberrations cause information loss, discontinuities, blurring and fragmentation, whereas FP-ExAO restores image quality, with cascaded DMs providing more uniform correction across the field than the pupil DM alone. **h-j,** XZ cross sections from the positions indicated in e-g under FP-ExAO (cascaded DMs), FP-ExAO (pupil DM) and NoAO conditions. Subpanels 1, 2 and 3 correspond to FP-ExAO (cascaded DMs), FP-ExAO (pupil DM) and NoAO, respectively. Colour coding indicates axial positions over a 1.5 μm range from -800 to 700 nm. Aberrations have a stronger effect along the z axis, while FP-ExAO restores the hollow outer mitochondrial membrane structures that are otherwise obscured. **k,** Quantitative summaries of SMLM localizations under FP-ExAO (cascaded DMs), FP-ExAO (pupil DM) and NoAO conditions. k1, distribution of localizations along the Z axis; k2, photon number distribution of localizations; k3, localizations per frame during SMLM acquisition. FP-ExAO recovers both the localization number and photon counts reduced by aberrations. Localizations and rendering were performed using SMAP. Scale bars, 20 μm in b-d, 10 μm in e-g and 1,000 nm in h-j.

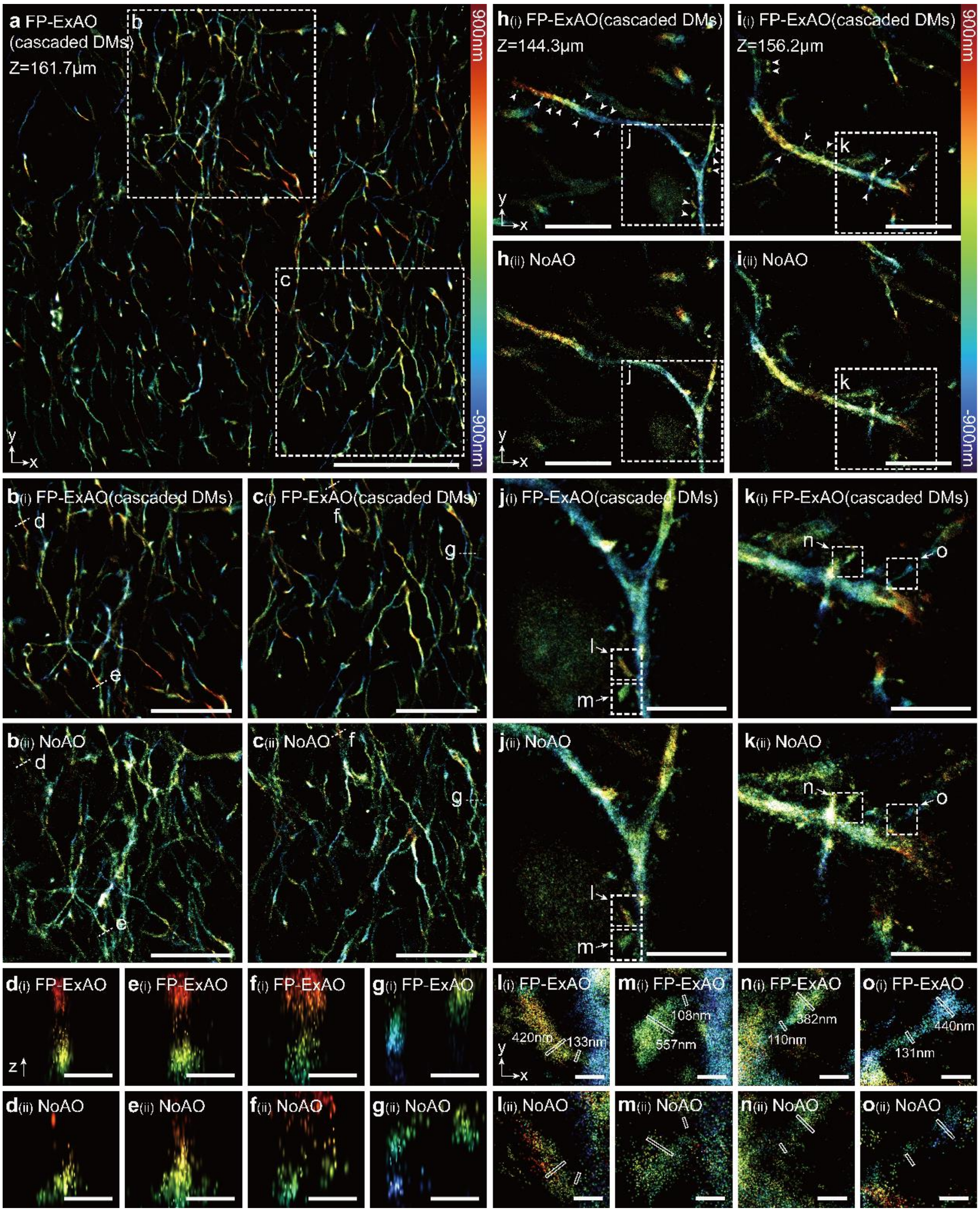

**Figure 4. Deep 3D SMLM imaging of neuronal structures in Thy1-EGFP mouse brain slices. a,** FP-ExAO (cascaded DMs) corrected 3D SMLM reconstruction of neuronal processes, mainly axons, in a cortical Thy1-EGFP mouse brain slice at an imaging depth of 161.7 μm. **b,c,** Magnified views of the two boxed regions in a. Subpanels i and ii show FP-ExAO (cascaded DMs) and NoAO reconstructions, respectively. FP-ExAO (cascaded DMs) increases image sharpness, axial information, structural continuity and localization recovery relative to NoAO. **d-g**, XZ cross sections at the positions indicated in b and c. Subpanels i and ii show FP-ExAO (cascaded DMs) and NoAO, respectively, demonstrating clearer separation of adjacent or crossing

structures along the optical axis after correction. **h,i,** 3D SMLM reconstructions of dendrites and dendritic spines at imaging depths of 144.3 µm (h) and 156.2 µm (i). Subpanels i and ii show FP-ExAO (cascaded DMs) and NoAO reconstructions, respectively. **j-o,** Magnified dendritic spine regions indicated in h and i. Subpanels i and ii show FP-ExAO (cascaded DMs) and NoAO, respectively. Width measurements in the FP-ExAO (cascaded DMs) images give spine neck widths of approximately 110–130 nm and spine head widths of approximately 300–600 nm in the displayed examples; corresponding structures are poorly discernible without correction in several regions. Colour coding indicates axial position over a 1.8 µm range from -900 to 900 nm. Localizations and rendering were performed using SMAP. Scale bars, 20 µm in a, 10 µm in b-c, h-i, 5 µm in j-k and 500 nm in d-g, l-o.

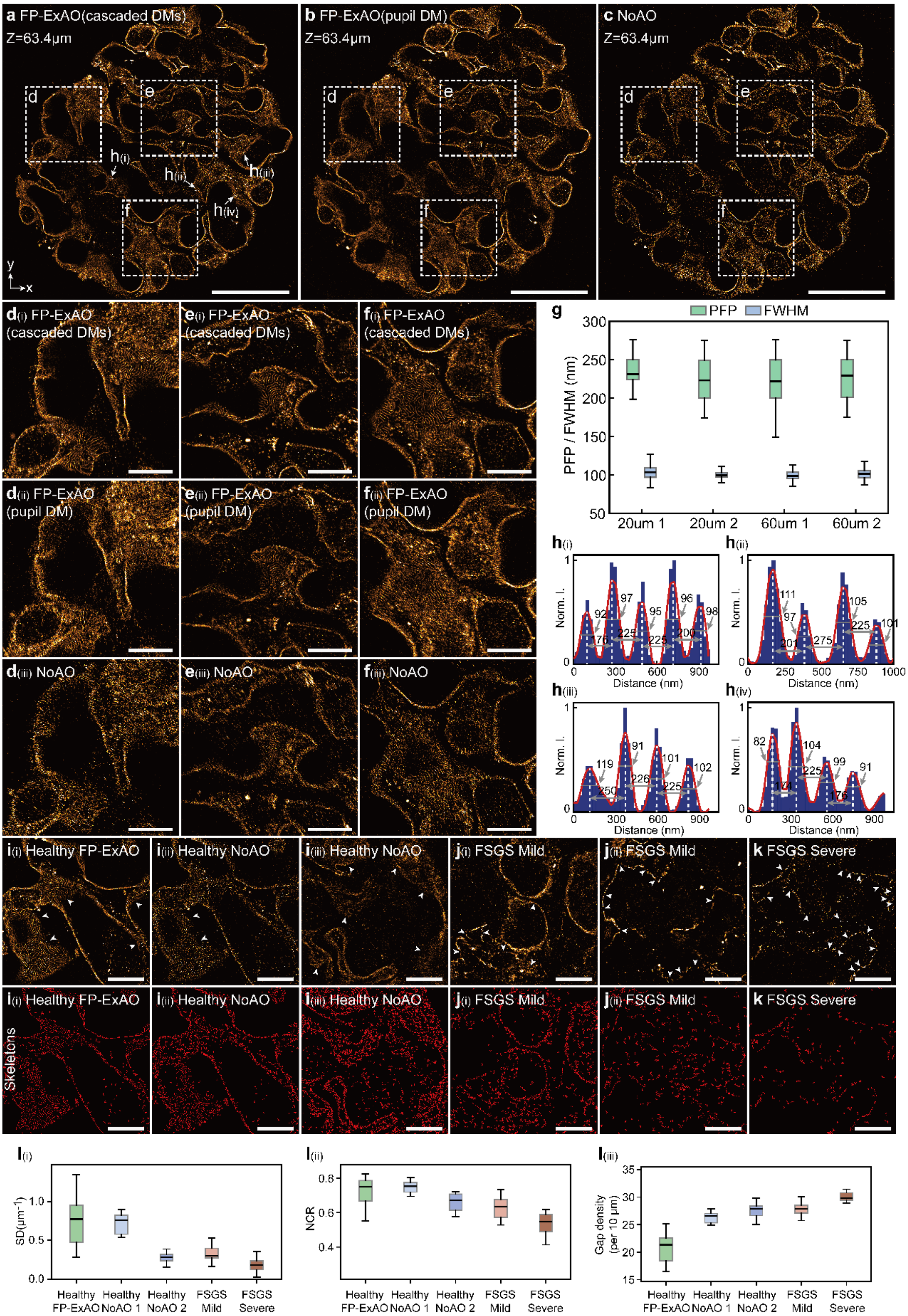

**Figure 5. Deep SMLM imaging of nephrin-labelled glomerular structures in mouse kidney tissue**. **a-c,** 3D SMLM reconstructions of nephrin-labelled structures in a mouse kidney section at an imaging depth of 63.4 μm under FP-ExAO (cascaded DMs) (a), FP-ExAO (pupil DM) (b) and NoAO (c) conditions. **d-f,** Magnified views

from field edge and field center regions comparing FP-ExAO (cascaded DMs), FP-ExAO (pupil DM) and NoAO. FP-ExAO (cascaded DMs) preserves the meandering nephrin-positive pattern across the field, whereas FP-ExAO (pupil DM) provides only partial correction in the presence of spatially varying aberrations and NoAO substantially reduces the visibility and continuity of the same structures. **g,h,** Quantification and representative line profiles of PFP structures and meandering nephrin-positive features. Most PFP measurements fall within the 150–300 nm range, consistent with the expected scale of these structures. **i,** Healthy mouse glomeruli imaged with FP-ExAO (cascaded DMs) and NoAO, together with the corresponding skeletonized nephrin-positive slit diaphragm networks. **j,k,** Nephrin-labelled glomerular structures and skeletonized networks in mild FSGS (j) and severe FSGS (k) samples imaged with FP-ExAO (cascaded DMs). **l,** Quantification of SD, NCR and gap density across Healthy FP-ExAO, Healthy NoAO 1, Healthy NoAO 2, FSGS Mild and FSGS Severe groups, based on images acquired from multiple sections per group, shows progressive loss and fragmentation of nephrin-positive slit diaphragm networks across the FSGS severity series, consistent with the morphological interpretation described in the manuscript. Localizations and rendering were performed using SMAP. Scale bars, 20 μm in a-c and 5 μm in d-f and i-k.

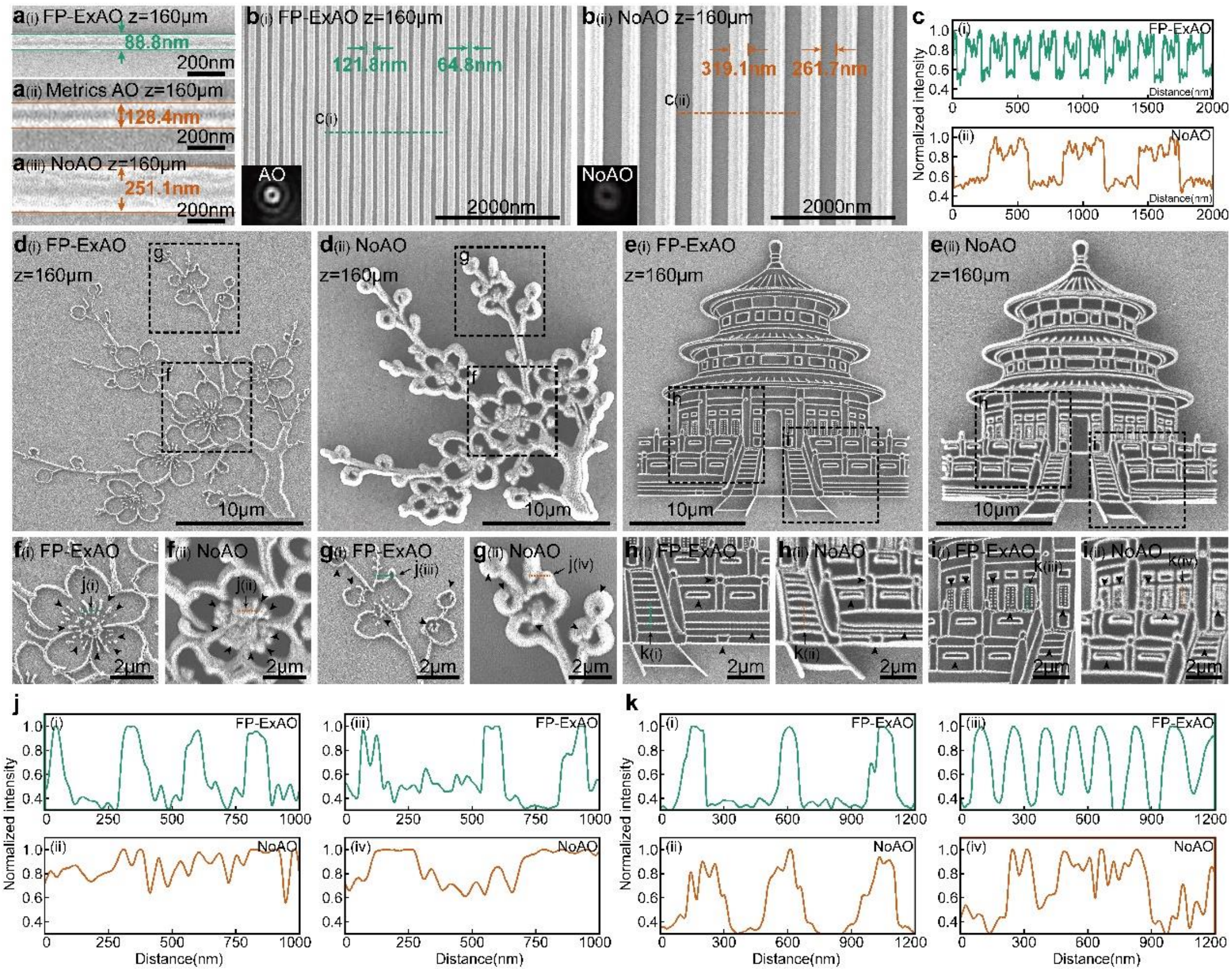

**Figure 6. SEM characterization of deep STED nanolithography with FP-ExAO. a,** Scanning electron microscopy (SEM) images of isolated lines fabricated at a writing depth of 160 μm with FP-ExAO correction (i), with Metrics-based AO correction (ii) and without FP-ExAO correction (NoAO, iii), with measured linewidths of 88.8 nm, 128.4 nm and 251.1 nm, respectively. **b,** SEM images of line gratings fabricated at the same depth under FP-ExAO (i) and NoAO (ii) conditions. Representative linewidths and inter line gaps are indicated in the images. Insets show the measured vortex depletion foci under the corresponding FP-ExAO and NoAO conditions. Dashed lines indicate the positions of the profiles shown in c. FP-ExAO reduces the minimum achievable feature size and outperforms Metrics AO. **c,** Normalized intensity profiles across the gratings in b. **d,** SEM images of plum blossom patterns fabricated under FP-ExAO (i) and NoAO (ii) conditions. Dashed boxes indicate the regions magnified in f and g. **e,** SEM images of Temple of Heaven patterns fabricated under FP-ExAO (i) and NoAO (ii) conditions. Dashed boxes indicate the regions magnified in h and i. FP-ExAO enables finer patterning of complex structures, particularly in details such as petals, stamens, stairs and windows. **f,g,** Magnified views of the central flower region (f) and peripheral branch region (g) from the plum blossom patterns in d. Arrowheads indicate representative fine structures, and coloured dashed lines indicate the positions of the profiles shown in j. **h,i,** Magnified views of representative window and step

regions from the Temple of Heaven patterns in e. Arrowheads indicate representative structural features, and coloured dashed lines indicate the positions of the profiles shown in k. **j,** Normalized intensity profiles measured from the positions indicated in f and g under FP-ExAO and NoAO conditions. **k,** Normalized intensity profiles measured from the positions indicated in h and i under FP-ExAO and NoAO conditions. Scale bars, 200 nm in a; 2 µm in b and f–i; and 10 µm in d and e.

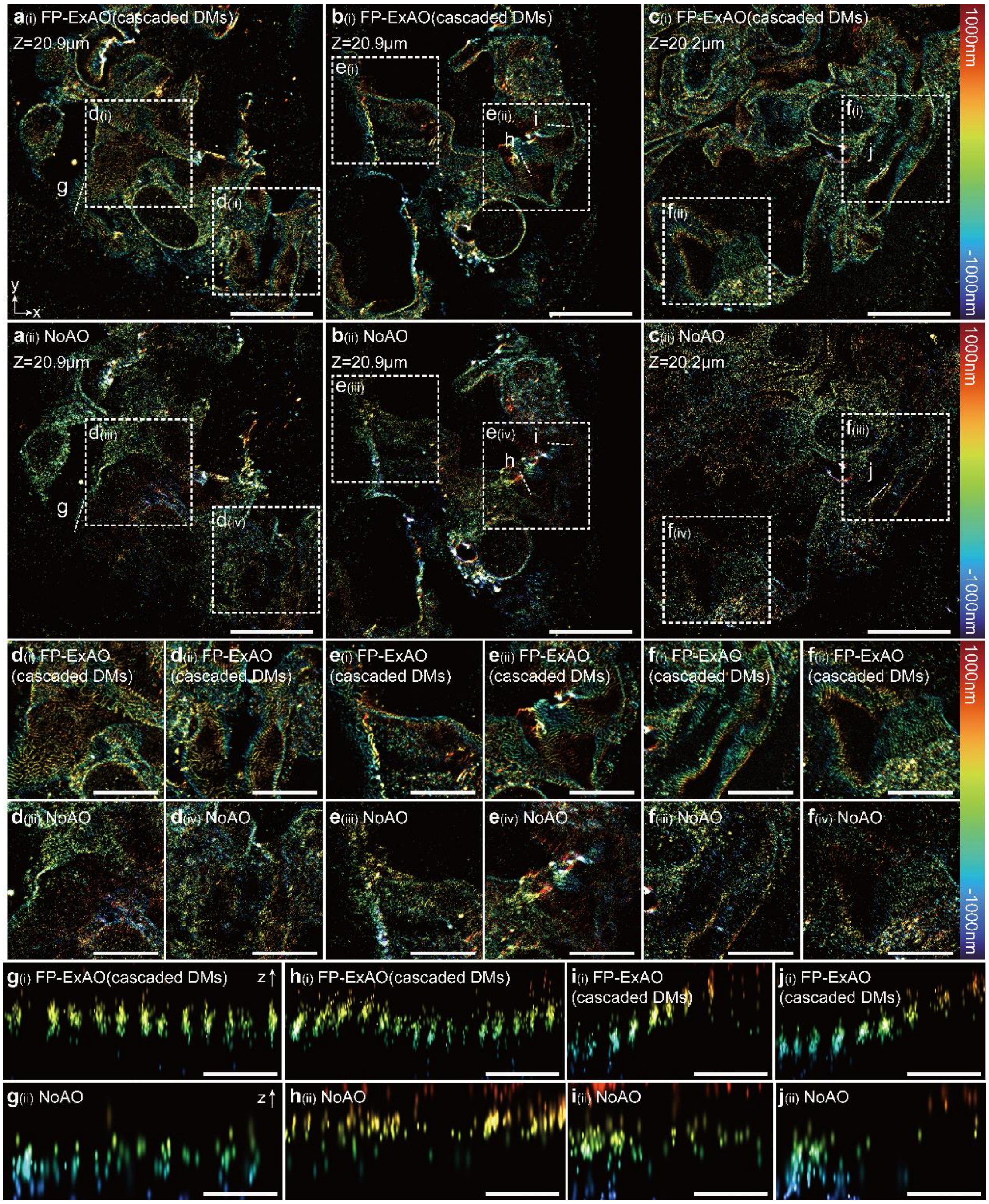

**Extended Figure 1. 3D SMLM imaging of nephrin-labelled structures in mouse kidney tissue with and without FP-ExAO. a-c,** 3D SMLM reconstructions of nephrin-labelled structures in mouse kidney sections at imaging depths of Z = 20.9 μm and Z = 20.2 μm, shown with FP-ExAO correction and without FP-ExAO correction. Panels a and b show representative regions at Z = 20.9 μm, and panel c shows a representative region at Z = 20.2 μm. Dashed boxes indicate the regions shown in d-f, and dashed lines indicate the positions of the XZ cross sections shown in g-j. **d-f,** Magnified views of the boxed regions in a-c comparing nephrin structures under FP-ExAO and NoAO conditions. FP-ExAO restores richer and more continuous nephrin

structures that are partially lost or blurred without correction. **g-j,** XZ cross sections from the positions indicated in a-c, comparing FP-ExAO and NoAO conditions. FP-ExAO more effectively recovers axial structural information and improves separation of nephrin features along the z axis. Colour coding indicates axial positions over a 2 µm axial range from -1,000 to 1,000 nm. Localizations and rendering were performed using SMAP. Scale bars, 10 µm in a-c, 5 µm in d-f and 1,000 nm in g-j.

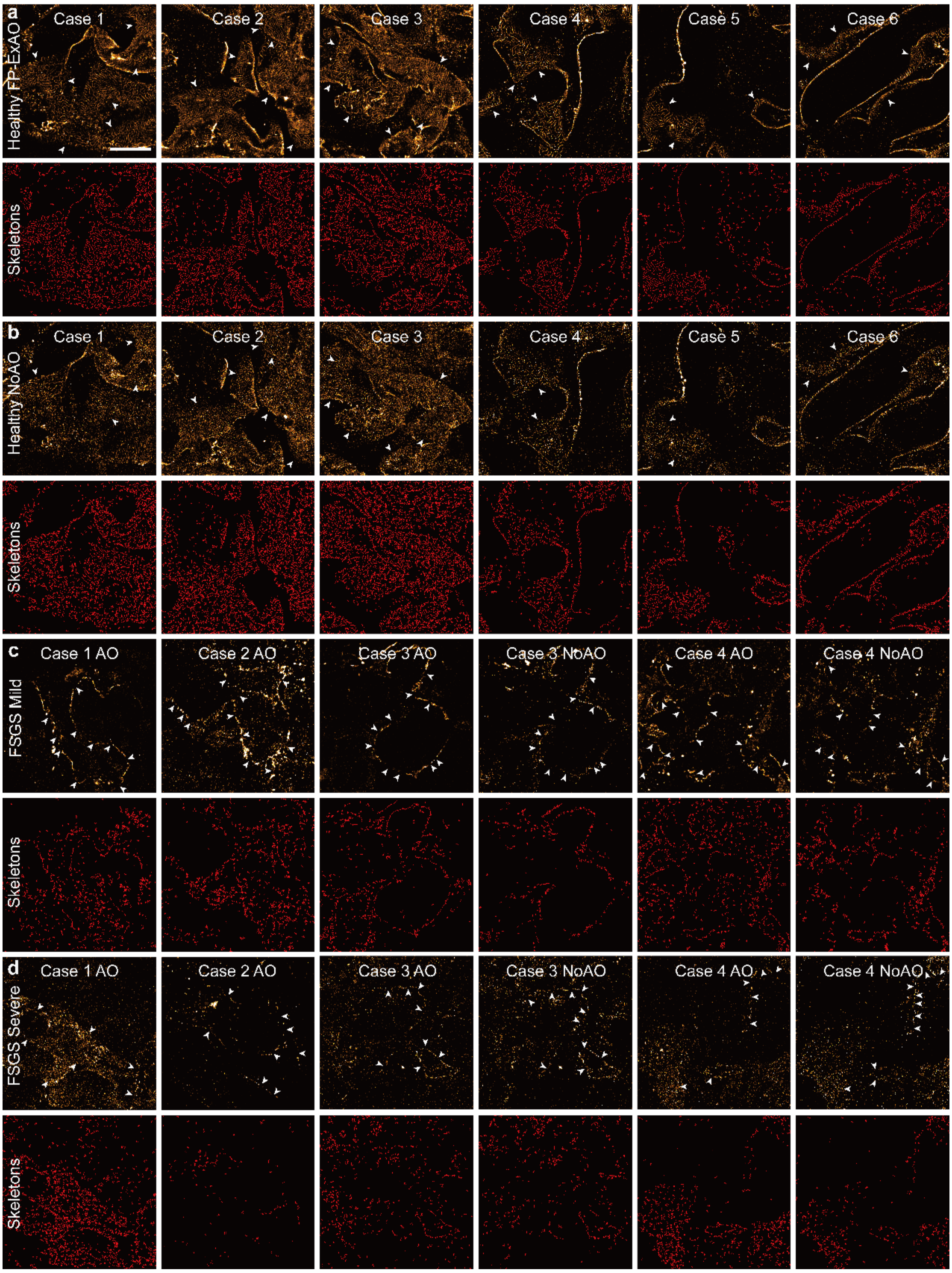

**Extended Figure 2. Additional examples of nephrin-labelled glomerular structures in healthy and FSGS kidney tissue. a,** Additional examples of healthy glomeruli imaged under FP-ExAO, together with corresponding skeletonized nephrin-positive slit diaphragm networks. **b,** Additional examples of healthy glomeruli imaged without AO correction and their corresponding skeletonized networks. **c,** Additional examples of mild FSGS glomeruli imaged under FP-ExAO and NoAO and their corresponding skeletonized networks. **d,** Additional examples of severe FSGS glomeruli imaged under

FP-ExAO and NoAO and their corresponding skeletonized networks. Six representative cases are shown for each group. The extended examples support the morphological trend shown in the main figure: nephrin-positive slit diaphragm networks are dense and continuous in healthy FP-ExAO images, appear less continuous without AO correction and show progressive loss and fragmentation in mild and severe FSGS. Scale bars, 5 μm.

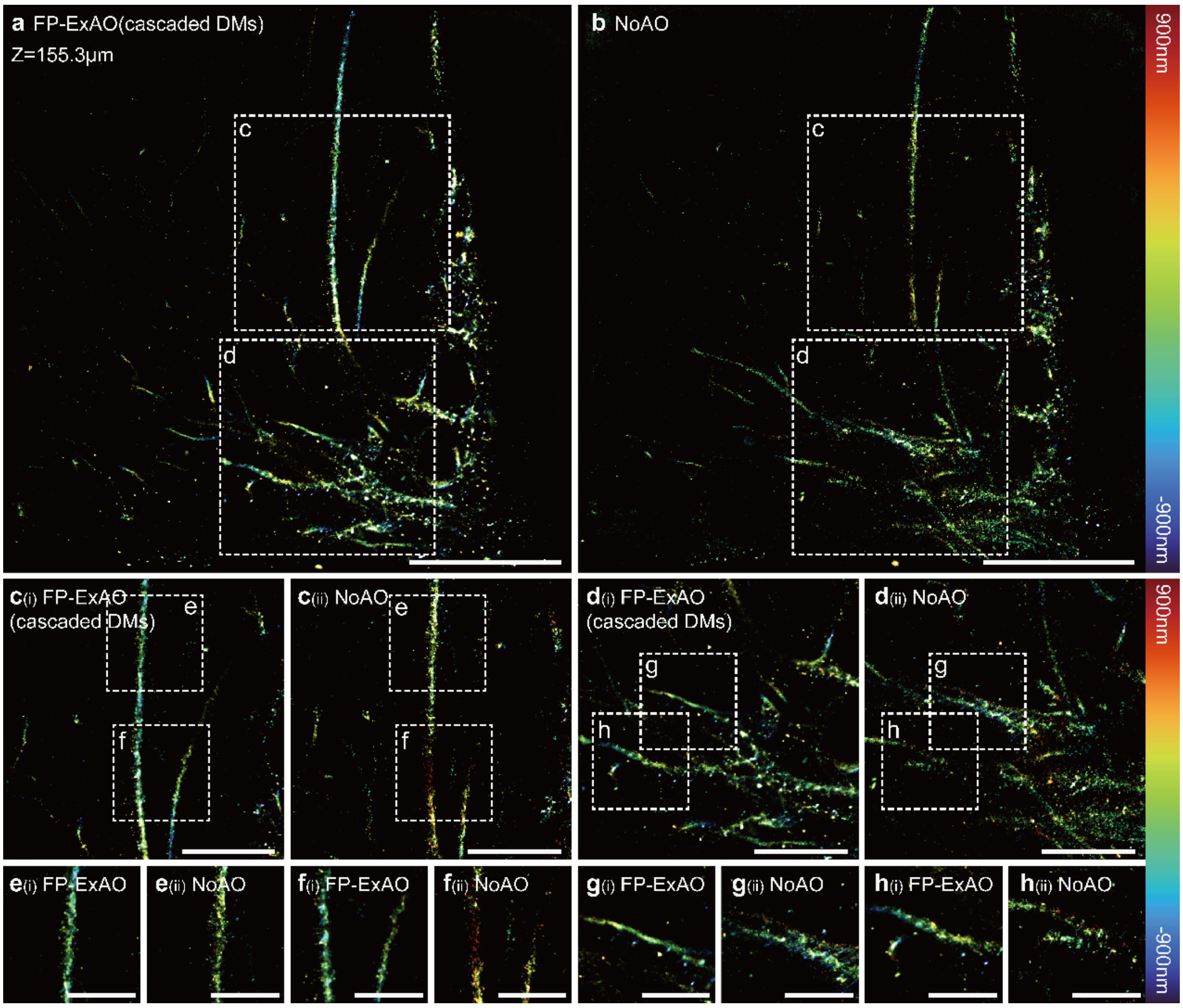

**Extended Figure 3. 3D SMLM imaging of GFAP-labelled astrocytic fibres in mouse brain tissue. a,b,** 3D SMLM reconstructions of GFAP-labelled astrocytic fibres in a mouse brain slice at an imaging depth of 155.3 μm under FP-ExAO (cascaded DMs) (a) and NoAO (b) conditions. Colour coding indicates axial position over a 1.8 μm range from -900 to 900 nm. **c,d,** Magnified views from the regions indicated in a and b, comparing GFAP-labelled structures with and without FP-ExAO (cascaded DMs) correction. **e-h,** Further magnified representative regions showing the recovery of fine GFAP-labelled fibre structures under FP-ExAO (cascaded DMs). FP-ExAO (cascaded DMs) increases localization recovery and structural continuity, whereas NoAO images contain more blurred or discontinuous fibres. Localizations and rendering were performed using SMAP. Scale bars, 20 μm in a-b, 10 μm in c-d and 5 μm in e-h.

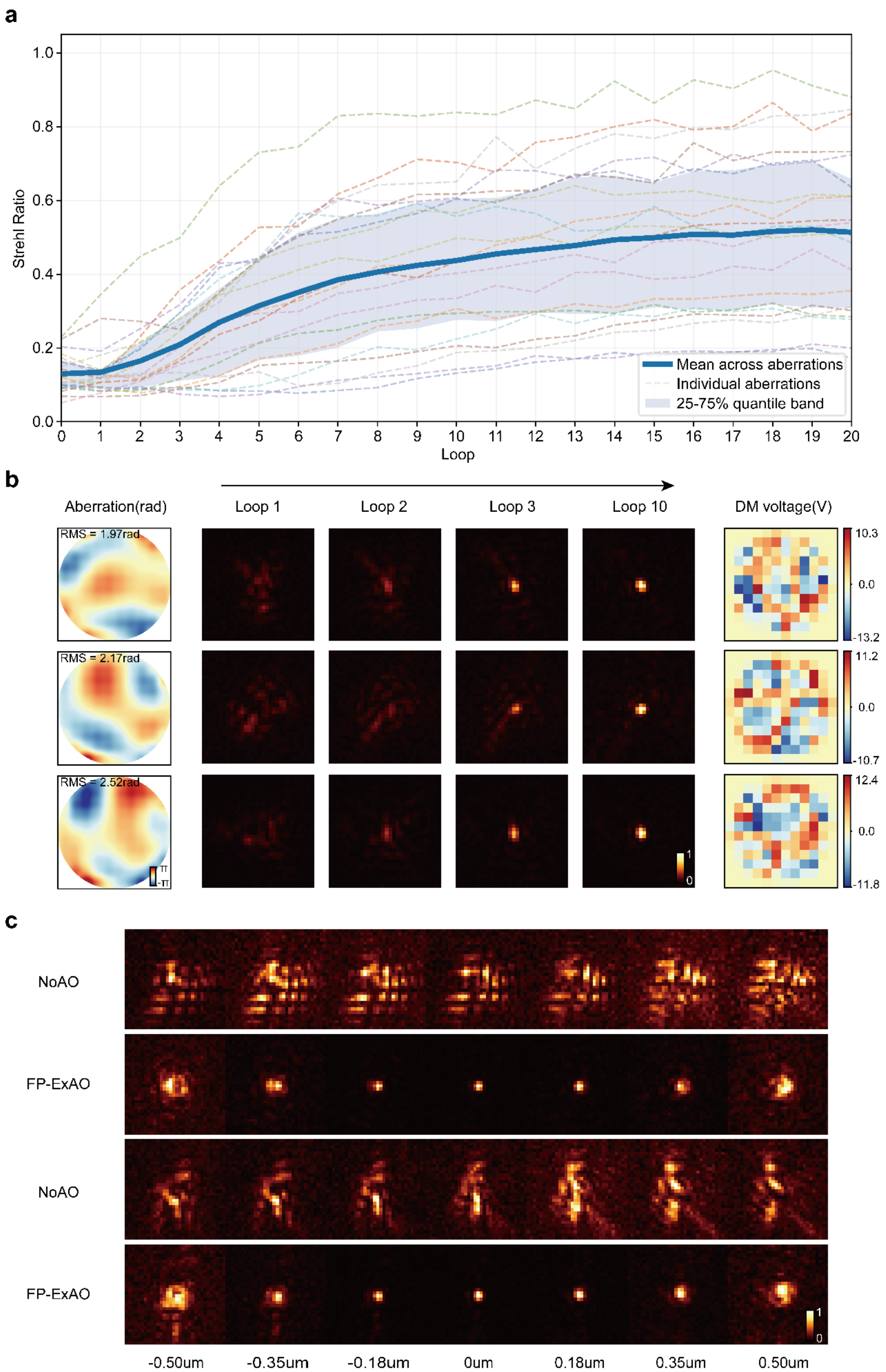

**Extended Figure 4. Simulation performance of FP-ExAO (pupil DM). a,** Closed-loop correction process of FP-ExAO under multiple randomly generated aberrations.

The Strehl ratio was used to quantify the correction process, with a Strehl ratio of 1 corresponding to the ideal aberration free focus. Light dashed curves indicate individual aberration tests, the blue solid curve indicates the mean across all tests, and the light blue shaded region indicates the interquartile range between the 25th and 75th percentiles. **b,** Representative correction examples under three artificially introduced aberrations. The left column shows the introduced phase aberrations, with the labelled values indicating the RMS phase errors; the middle columns show PSF images at correction loops 1, 2, 3 and 10; and the right column shows the final deformable mirror voltage maps. FP-ExAO progressively concentrates the focal energy during closed-loop correction and converges reliably under different random aberrations. **c,** Axial comparison of 3D PSF image sequences before and after FP-ExAO correction. The two rows labelled NoAO and FP-ExAO show PSFs at the indicated axial positions from -0.50 μm to 0.50 μm. FP-ExAO restores the 3D focal structure and improves axial confinement of the PSF.

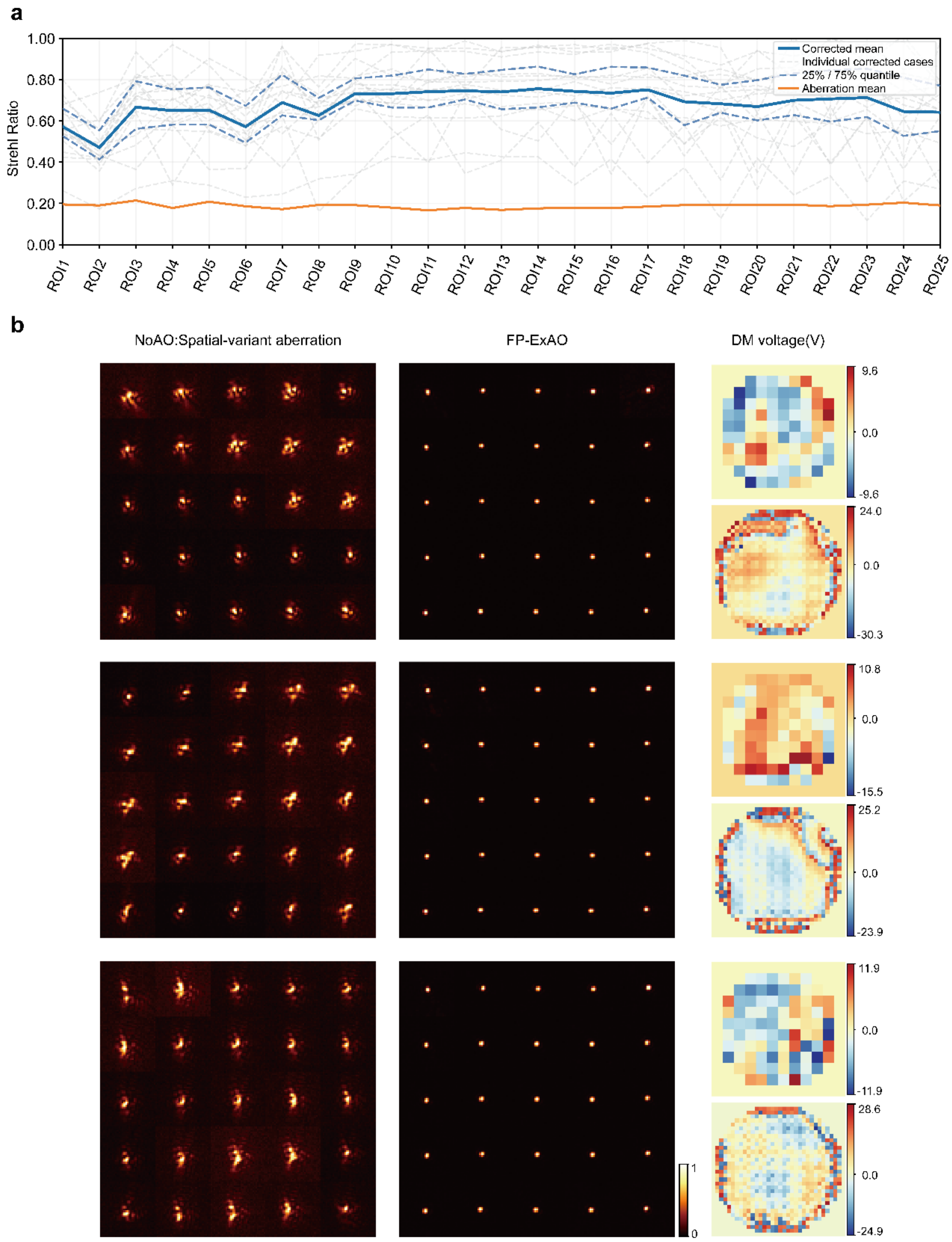

**Extended Figure 5. Simulation performance of FP-ExAO for spatially varying aberration correction. a,** Full field correction results of FP-ExAO under artificially introduced spatially varying aberrations. The field of view was divided into 5 × 5 isoplanatic regions, corresponding to ROI 1-ROI 25. The vertical axis shows the Strehl ratio for each ROI. The blue solid curve indicates the corrected mean, light grey dashed curves indicate individual corrected cases, blue dashed curves indicate the 25th and 75th percentiles, and the orange curve indicates the aberration mean. **b,** Representative

PSF arrays before and after FP-ExAO correction for three spatially varying aberrations, together with the corresponding deformable mirror voltage maps. The left column shows PSF arrays under spatially varying aberration, the middle column shows PSF arrays after FP-ExAO correction, and the right column shows the corresponding deformable mirror voltage maps with voltage colour scales. These results demonstrate that FP-ExAO enables cascaded DMs to correct spatially varying aberrations, providing uniform correction across the field.

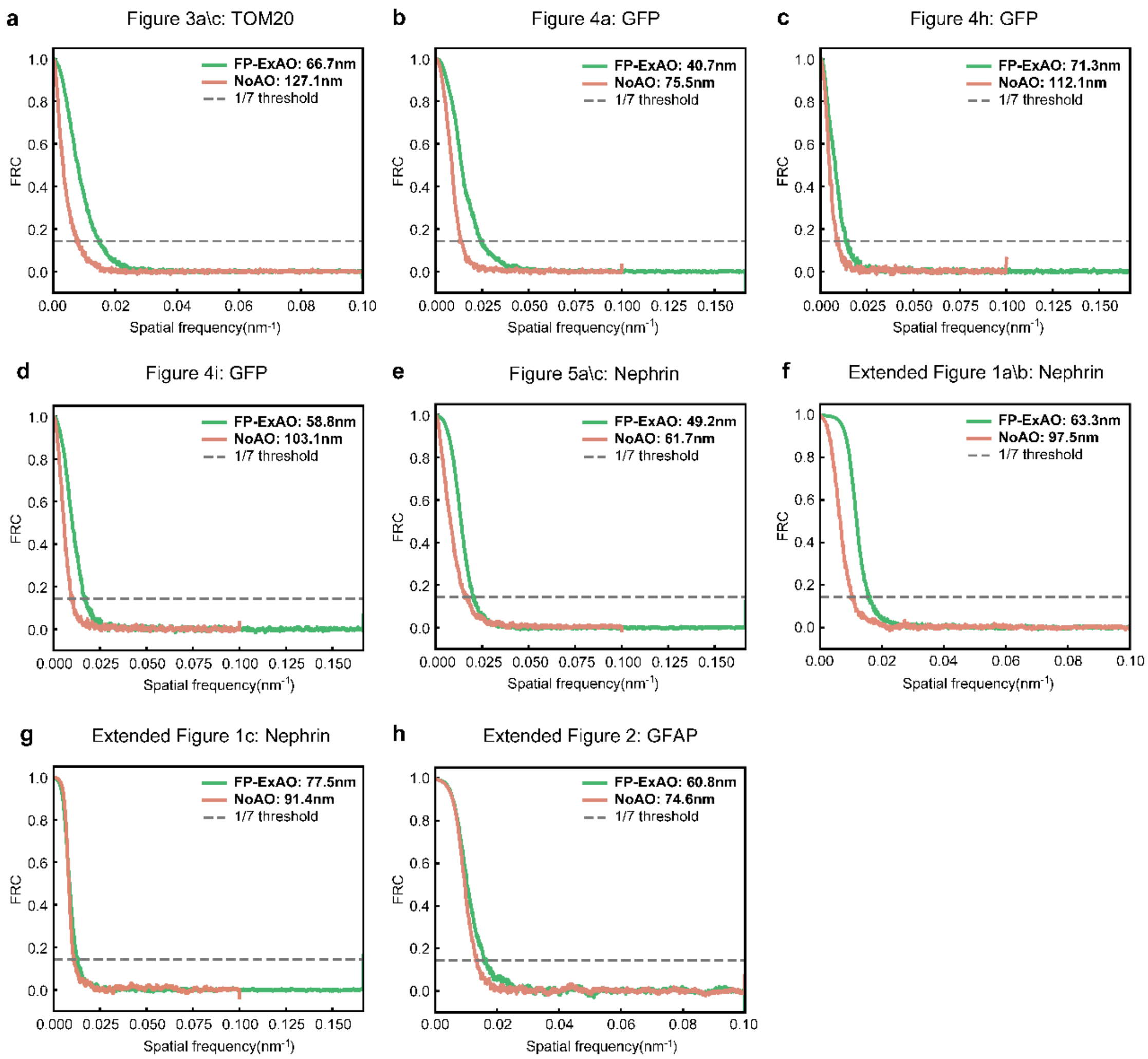

**Extended Figure 6. FRC resolution analysis of SMLM imaging results with and without AO.** FRC resolution was evaluated using the 1/7 threshold criterion. **a,** FRC analysis of TOM20 shown in Fig. 3b,d, with resolutions of 66.7 nm for FP-ExAO (cascaded DMs) and 127.1 nm for NoAO. **b,** FRC analysis of GFP-labelled neuronal structures shown in Fig. 4a, with resolutions of 40.7 nm for FP-ExAO (cascaded DMs) and 75.5 nm for NoAO. **c,** FRC analysis of GFP-labelled neuronal structures shown in Fig. 4h, with resolutions of 71.3 nm for FP-ExAO (cascaded DMs) and 112.1 nm for NoAO. **d,** FRC analysis of GFP-labelled neuronal structures shown in Fig. 4i, with resolutions of 58.8 nm for FP-ExAO (cascaded DMs) and 103.1 nm for NoAO. **e,** FRC analysis of nephrin-labelled kidney structures shown in Fig. 5a,c, with resolutions of 49.2 nm for FP-ExAO (cascaded DMs) and 61.7 nm for NoAO. **f,** FRC analysis of nephrin-labelled kidney structures shown in Extended Fig. 1 a,b, with resolutions of 63.3 nm for FP-ExAO (cascaded DMs) and 97.5 nm for NoAO. **g,** FRC analysis of nephrin-labelled kidney structures shown in Extended Fig. 1c, with resolutions of 77.5 nm for FP-ExAO (cascaded DMs) and 91.4 nm for NoAO. **h,** FRC analysis of GFAP-labelled structures shown in Extended Fig. 3, with resolutions of 60.8 nm for FP-ExAO (cascaded DMs) and 74.6 nm for NoAO.

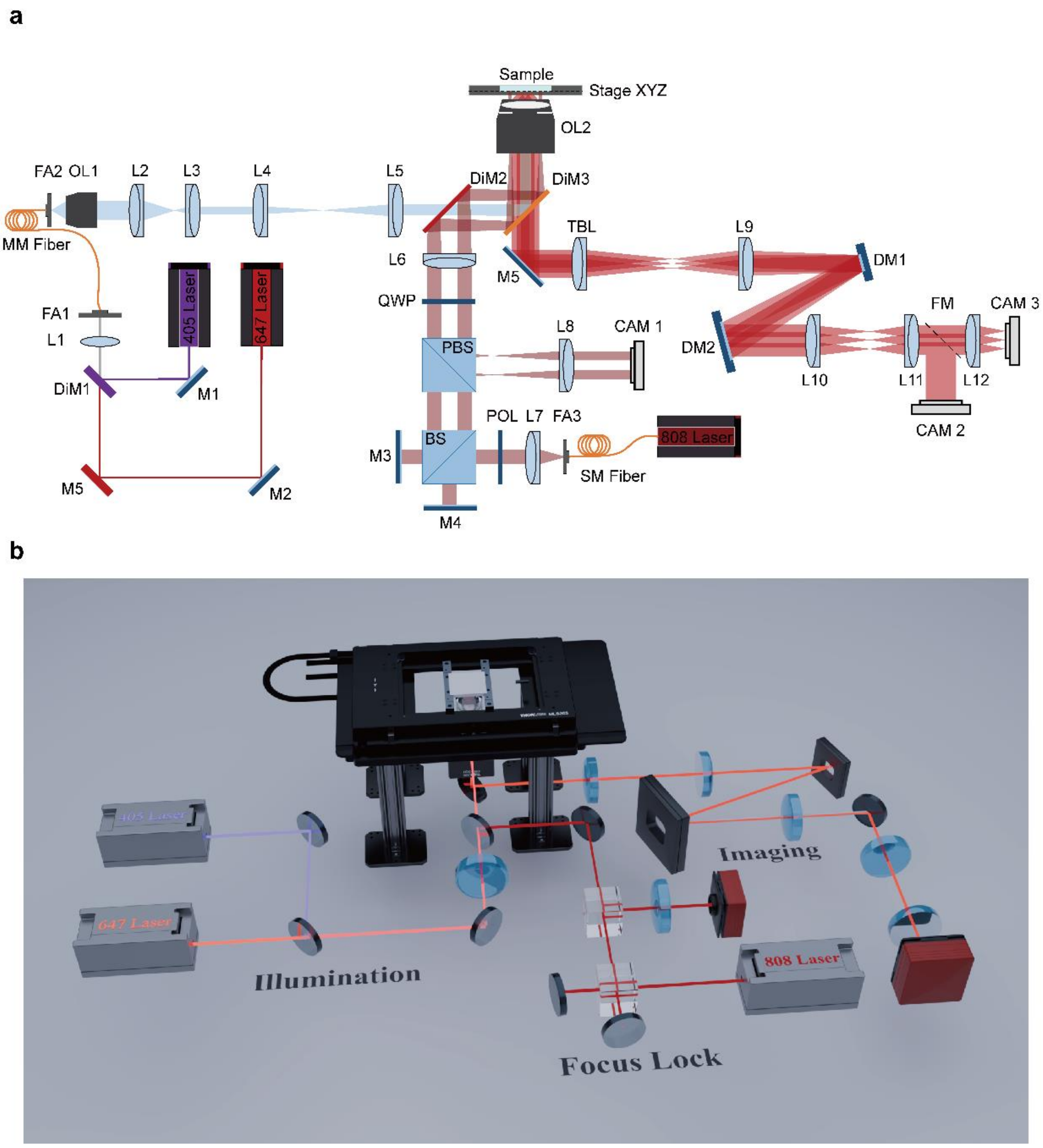

**Extended Figure 7. Optical and mechanical layout of the FP-ExAO SMLM microscope. a,** Optical path of the FP-ExAO SMLM microscope, including the illumination module, imaging and AO module, and focus lock module. **b,** Rendered 3D mechanical layout of the microscope, showing the spatial arrangement of the illumination, focus lock and imaging modules. M, mirror; DiM, dichroic mirror; FM, flip mirror; L, lens; OL, objective lens; FA, fibre adapter; MM fibre, multimode fibre; SM fibre, single mode fibre; BS, beam splitter; PBS, polarizing beam splitter; QWP, quarter wave plate; POL, polarizer; TBL, tube lens; DM, deformable mirror; CAM, camera; Stage XYZ, motorized three axis sample stage. L2-L12 are achromatic lenses.

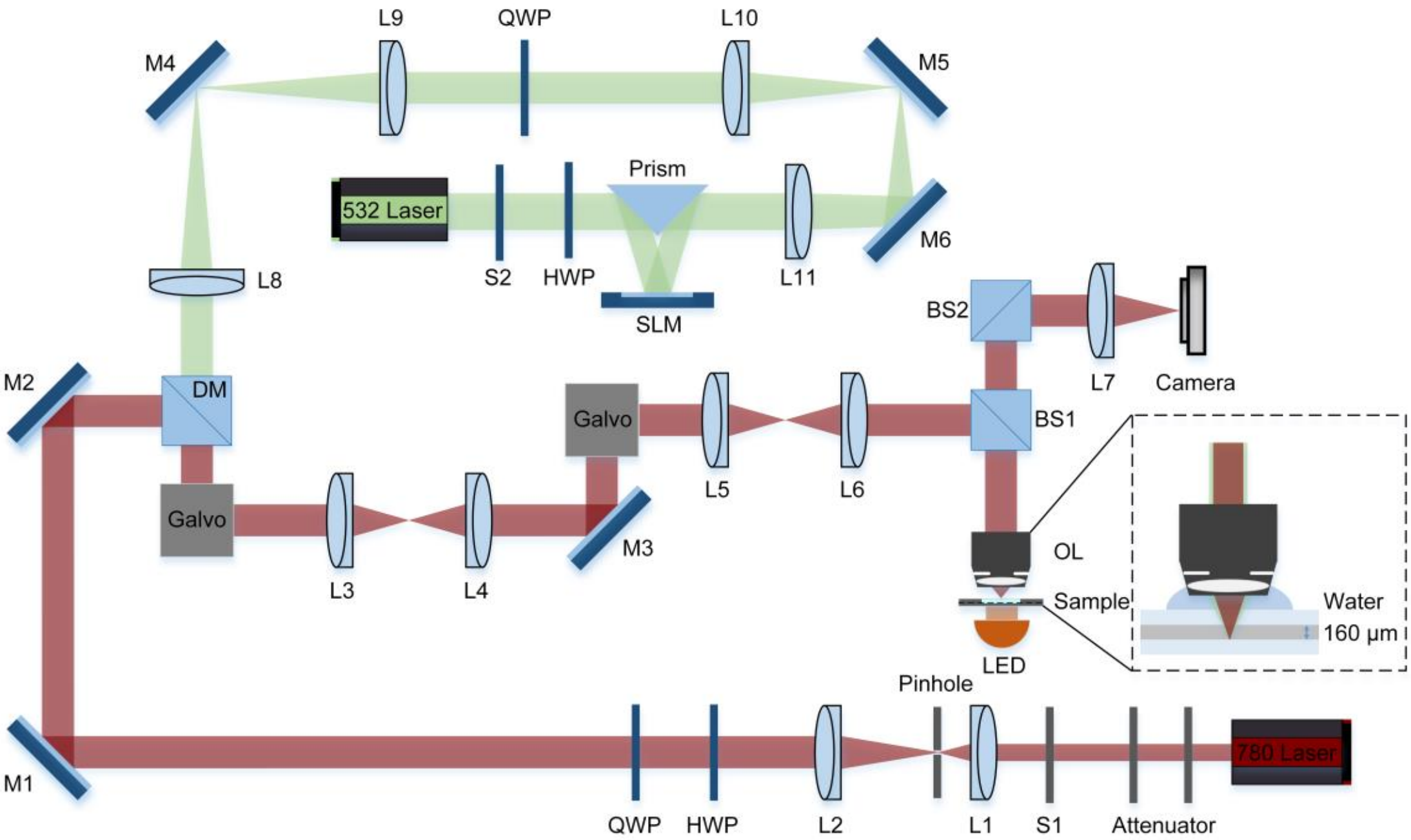

**Extended Figure 8. Optical layout of the STED direct laser writing system for nanolithography.** Optical layout of the STED direct laser writing system, including the excitation module, vortex depletion module, beam scanning module, and in situ transmission monitoring module. M, mirror; S, shutter; HWP, half wave plate; QWP, quarter wave plate; SLM, spatial light modulator; L, lens; DM, dichroic mirror; BS, beam splitter; OL, objective lens; CAM, camera.